\documentclass[PRB,reprint,amsmath,amssymb,longbibliography,superscriptaddress]{revtex4-1}
\usepackage{graphicx}
\usepackage{dcolumn}
\usepackage{bm}
\usepackage{hyperref}
\usepackage{breakurl}
\usepackage{multirow}
\usepackage{enumitem}
\usepackage[pagewise]{lineno}
\setcitestyle{super}

\begin{document}

\title{Exploring Topological Superconductivity in Topological Materials}

\author{Yupeng Li}
      \affiliation{Zhejiang Province Key Laboratory of Quantum Technology and Device, Department of Physics, Zhejiang University, Hangzhou 310027, China}

\author{Zhu-An Xu}
      \email{zhuan@zju.edu.cn}
      \affiliation{Zhejiang Province Key Laboratory of Quantum Technology and Device, Department of Physics, Zhejiang University, Hangzhou 310027, China}
      \affiliation{Zhejiang California International NanoSystems Institute, Zhejiang University, Hangzhou 310058, China}
      \affiliation{Collaborative Innovation Centre of Advanced Microstructures, Nanjing University, Nanjing 210093, China}

\date{\today}

\begin{abstract}
The exploration of topological superconductivity and Majorana zero modes has
become a rapidly developing field. Many types of
proposals to realize topological superconductors have been presented, and significant
advances have been recently made. In this review, we conduct a survey
on the experimental progress in possible topological
superconductors and induced superconductivity in topological
insulators or semimetals as well as artificial structures.
The approaches to inducing superconductivity in topological
materials mainly include high pressure application, the hard-tip point contact method,
chemical doping or intercalation, the use of artificial topological
superconductors, and electric field gating. The evidence
supporting topological superconductivity and signatures of
Majorana zero modes are also discussed and summarized.

\end{abstract}

\maketitle

\section*{Introduction}
Topological superconductors (TSCs) with protected Majorana surface
states have become a focus of interest due to their potential
applications in the fault-tolerant topological quantum computation.\cite{topologicalquantumcomput_Nayak_RMP08}
Topological classifications of gapped quantum many-body systems
have sparked a revolution in condensed matter physics.
Nontrivial topology in relation to superfluid helium 3 ($^{3}$He) was discussed by
Volovik in 2003,\cite{HeliumDroplet_Volovik_Oxford03} and
topologically nontrivial superconductors were introduced approximately two
decades ago in a two-dimensional (2D) model by Read and Green
\cite{Paired_ReadN_PRB00} as well as in a one-dimensional (1D) model by
Kitaev.\cite{quantumwire_Kitaev_pu01} A non-Abelian Majorana zero
mode in the vortex core in the 2D case or at the edge in the 1D
case was theoretically predicted in these works, long before the
recent research boom in the field of topological phases of matter.

Majorana fermions, whose particles are their own
antiparticles, were first introduced by Ettore Majorana in the
context of elementary particle physics.\cite{MajoranaF_Majorana_Il1937}
Unlike in particle physics, in condensed matter
physics, if the creation operator $\gamma^+$ of a quasiparticle, which is a
superposition of electron and hole excitations, is the same as
the annihilation operator $\gamma$, then this type of a particle can be
regarded as a Majorana fermion. In the Read-Green model,\cite{Paired_ReadN_PRB00}
the Bogoliubov quasiparticles (neutral
excitations in superconductors) in the bulk can be identified as dispersive
Majorana fermions, which can be bound to a defect at zero energy,
and these combined objects are called Majorana bound states or Majorana zero modes (MZMs).\cite{Majorana_Wilczek_NatPhys,review_Beenakker_ARCMP13}
MZMs localized at defects have been theoretically predicted to obey exotic exchange
statistics, similar to non-Abelian anyons, which will change their
quantum states if they are interchanged. This phenomenon is the most attractive
property of MZMs. Exchanging the position of MZMs corresponds to a nontrivial
transformation within the degenerate ground-state manifold and
represents a noncommutative operation, which does not depend on
the method or details of the execution. Such an operation
is topologically protected and only depends on the exchange
statistics of quasiparticles; thus, it can be used in
quantum gates or topological quantum computing.

Unfortunately, little progress was made in realizing TSCs before the discovery of topological insulators
(TIs). Intrinsic TSCs are rare because they require
odd-parity pairing, and such pairing states usually require a
particular combination of pairing interactions and electronic
structure. A conceptual breakthrough in the context of TIs emerged from the
seminal work of Fu and Kane.\cite{TI3D_FuL_PRL07,TIs_Ful_PRB07}
Their theory predicts that surface Dirac fermions of
TIs may realize topological superconductivity when in proximity to an s-wave superconductor.\cite{SCpeonTI_FuL_PRL08,TIandTSC_QiXL_RMP11} Subsequently, Fu
and Berg further propose a new mechanism for odd-parity pairing
facilitated by strong spin-orbit coupling (SOC) such that
time-reversal-invariant TSCs can be realized with carrier-doped TIs
such as Cu$_{x}$Bi$_{2}$Se$_{3}$.\cite{CuxBi2Se3Theory_FuL_PRL10} Theoretical proposals on the realization of
MZMs in hybrid systems of s-wave superconductors deposited on the
surface of a strong TI and related structures,\cite{SCpeonTI_FuL_PRL08,semicSCnewp_SauJD_PRL10,review_Beenakker_ARCMP13}
such as semiconducting nanowires, have motivated intense efforts to `engineer'
topological superconducting states through the simple building of components.

Strong TSCs have three characteristics:\cite{TIandTSC_QiXL_RMP11}
(1) odd-parity pairing symmetry with a full superconducting gap;
(2) gapless surface states consisting of Majorana fermions; and (3)
MZMs in the superconducting vortex cores. In addition, unconventional
superconductors usually have nodal superconducting gaps. If the
nodes themselves have topological properties, such unconventional
superconductors can also be regarded as TSCs, and they are called
weak TSCs.\cite{TSC_Sato_RPP17}

Generally, TSCs can also be classified as time-reversal-breaking TSCs and time-reversal-invariant TSCs according to the presence or absence of internal symmetries (such as time reversal symmetry and spin rotation symmetry),\cite{TCIandTSC_Ando_ARCMP15} respectively. Sr$_{2}$RuO$_{4}$ is often
regarded as a typical example of the former type. The existence of an internal magnetic
field in the superconducting state that breaks the time reversal symmetry
has been supported by nuclear magnetic resonance (NMR) measurements
\cite{Sr2RuO4NMR_Ishida_Nature98} and muon spin rotation ($\mu$SR)
experiments.\cite{Sr2RuO4USR_Luke_Nature98,UPt3uSR_Luke_PRL93}
The other approach to achieving time-reversal-breaking TSCs is to
fabricate semiconductor/superconductor heterostructures in which
the MZMs are located at the ends of the semiconductor nanowire.\cite{InSbSC_Mourik_science12}
The latter type, as a new type of TCS, requires strong SOC to
induce odd-parity pairing. Cu$_{x}$Bi$_{2}$Se$_{3}$ is proposed
as a promising candidate for this type.\cite{CuxBi2Se3Theory_FuL_PRL10}
Moreover, the proximity effect in TI/superconductor heterostructures will
also induce time-reversal-invariant TCSs in which the MZMs could exist at
the vortex cores. Usually, MZMs can be characterized by the zero-bias peak
(ZBP) in the differential conductance obtained  through scanning tunneling
microscopy (STM) measurements.\cite{Bi2Te3-NbSe2MF_XuJP_PRL15,Bi2Te3-NbSe2_SunHH_PRL16}
Other features of these TCSs, such as  gapless surface
states, can be detected by measuring nontrivial
transport properties, for instance, a 4$\pi$-periodic Josephson supercurrent \cite{HgTe-NbMF_Wiedenm_NC16,HgTe-AlMF_Deacon_PRX17} and helical
edge states in the supercurrent.\cite{HgTe/HgCdTe_Hart_NP14}

A variety of routes can be followed to realize topological superconductivity, and
many experimental reports about confirming MZMs have been presented.
This review mainly focuses on the superconductivity induced in
materials with nontrivial topology as well as the superconductivity
arising through the proximity effect in `hybrid systems', and the current
situation of pristine TSCs is also briefly summarized. In
section I, we will first introduce the TSC candidates
among pristine superconductors, such as Sr$_{2}$RuO$_{4}$,
UPt$_{3}$, and iron-based superconductors with nontrivial
topological features. In section II, the high pressure technique is
introduced as a common method to realize
superconductivity in topological materials, such as TIs, Dirac
semimetals (DSMs), and Weyl semimetals (WSMs). However, performing further
experiments to observe the pairing states and gapless surface states under pressure
is quite difficult. In section III, the hard-tip
point contact method, similar to the high pressure technique, provides a
new platform to obtain superconductivity in topological
materials. One advantage of this technique is that the Andreev
reflection can be measured to detect MZMs. In
section IV, superconductivity induced in topological materials by
either doping or intercalation is summarized, which is certainly
worth further investigating. In section V, artificial
structures, such as nanowire/superconductor heterostructures and
TI/superconductor heterostructures,
are introduced, which are the most successful approaches to realizing
TSCs, and very strong signatures of MZMs are presented. Moreover, an electric field gating technique is also introduced because it is widely used in the measurement of artificial structures. Finally,
a summary and an outlook on this field are presented.

\section{Pristine topological superconductors}

To date, most superconductors are s-wave superconductors with
spin-singlet pairing, and such conventional s-wave pristine
superconductors are clearly nontopological because they exhibit a
smooth crossover from the weak-coupling Bardeen-Cooper-Schrieffer
(BCS) limit to the strong-coupling Bose-Einstein condensate (BEC) limit without undergoing a
gap-closing phase transition. Namely, we should search for TSCs by studying
superconductors with unconventional pairing symmetry. We know that the
total spin ($S$) of a Cooper pair has two states:\cite{superconductivity_JFAnnett_book04}
the spin-singlet pairing state with $S$ = 0 and the
spin-triplet pairing state with $S$ = 1. Therefore, odd parity
of the wave function of a Cooper pair ($S$ = 0) requires even parity of
the orbital wave function ($L$ = 0, 2, 4 ...),
satisfying opposite parity of the total wave function. Moreover,
$L$ = 0 or 2 corresponds to s-wave or d-wave pairing,
respectively. In spin-triplet states ($S$ = 1), the orbital wave function should be odd parity [$L$ = 1 (p-wave), 3 ($f$-wave), 5 ...].

In general, a majority of superconductors are known to exhibit
spin-singlet pairing states, such as conventional superconductors,
cuprate superconductors,\cite{CuprateSC_Tsuei_RMP00}
and iron-based superconductors.\cite{IronbasedSC_Stewart_RMP11}
Therefore, to search for TSCs, time-reversal-symmetry-breaking superconductors with spin-triplet pairing states
have attracted tremendous attention, although they are quite
rare. If pristine superconductors with unconventional pairing
symmetry have nontrivial topology, they can be called
`intrinsic' TSCs. A relatively small number of
candidates for spin-triplet superconductors exist. Several candidates, based on
substantial evidence, may possess spin-triplet pairing, but none has
been as firmly confirmed as superfluid $^{3}$He.
The promising candidates for spin-triplet pairing are
Sr$_{2}$RuO$_{4}$ and UPt$_3$, although their superconducting
transition temperature ($T_{c}$) is low. Other possible spin-triplet
superconductors include ferromagnetic heavy fermion
superconductors, such as UGe$_{2}$ and UCoGe, and
noncentrosymmetric (lacking inversion symmetry) superconductors,
such as Li$_{2}$Pt$_{3}$B.\cite{Sr2RuO4_MaenoY_JPCS12} The
recently discovered K$_{2}$Cr$_{3}$As$_{3}$ family with a
quasi-one-dimensional (quasi-1D) structure may also have an
unconventional pairing state.\cite{CrBasedSC_CaoGH_PM17} In the
following subsections, we will present the updated progress in
the Sr$_{2}$RuO$_{4}$, UPt$_3$ \cite{TSC_Sato_RPP17} and iron-based superconductors.
Regarding the K$_{2}$Cr$_{3}$As$_{3}$ family, little progress has been made
in terms of their pairing states due to their extremely air-sensitive properties.

\subsection{Sr$_{2}$RuO$_{4}$}

Stimulated by the discovery of the high $T_c$ cuprate superconductors
La$_{2-x}$Ba$_{x}$CuO$_{4}$,\cite{La2BaxCuO4_Bednorz_ZPB86} the
element substitution method has been widely employed for perovskite-type tetragonal
crystal structures to search for new high
$T_{c}$ superconductors, and a large family of high $T_c$ cuprate
superconductors has emerged. Many attempts to replace copper with other 3d
transition metals have been made, which finally led to the
discovery of a new interesting superconductor, Sr$_{2}$RuO$_{4}$,
with $T_{c} \approx$ 1 K in 1994.\cite{Sr2RuO4_Maeno_Nature94} This
new superconductor has several unusual properties, such as an
impurity-sensitive $T_c$, absence of a coherence peak in the NMR spectrum, and
a linear specific heat at low temperature.\cite{Sr2RuO4_Mack_RMP03}

The density of states at the Fermi surface in
normal metals can generate a paramagnetic susceptibility
$\chi_{p}$, i.e., the Pauli susceptibility, and a magnetic field ($B$)
can lower the free energy by $\frac{1}{2}\chi_{p}B^{2}$.\cite{PauliLimit_Clogston_PRL62}
However, superconductors with spin-singlet pairing states would suppress paramagnetic
susceptibility at absolute temperature because no electron exists
at the Fermi surface and Cooper pairs form instead. Therefore,
spin susceptibility disappears in the spin-singlet pairing
states. However, the spin-triplet state allows paramagnetic
susceptibility, and this characteristic can be employed as an
experimental clue to spin-triplet superconductivity. NMR is an
effective method to measure the spin states of electrons. The Knight shift
related to the spin susceptibility was reported to remain unchanged in
Sr$_{2}$RuO$_{4}$ when the temperature is below $T_{c}$.\cite{Sr2RuO4NMR_Ishida_Nature98}
In addition, the time-reversal-symmetry-breaking superconductors
exhibit an internal magnetic field below $T_{c}$ due to the spin-triplet
pairing states, which has also been successfully detected by $\mu$SR
experiments.\cite{Sr2RuO4USR_Luke_Nature98}

Another key experiment to support this spin-triplet state of
Sr$_{2}$RuO$_{4}$ is phase-sensitive measurements.\cite{Sr2RuO4PS_Nelson_Science04} For such studies,
a superconducting quantum interference device (SQUID) with a
Josephson junction of Au$_{0.5}$In$_{0.5}$-Sr$_{2}$RuO$_{4}$ located on the opposite
sides of the Sr$_{2}$RuO$_{4}$ crystal was designed and prepared.
The critical current $I_{c}$ across the Josephson junctions will
be minimized if no flux ($\Phi$ = 0) occurs in the SQUID according to
the following formula:\cite{Sr2RuO4PS_Nelson_Science04,IcJunction_Geshk_PRB87}
\begin{equation}\label{Ic}
I_{c} \propto cos\left[\left(\frac{\Phi}{\Phi_{0}}+\frac{1}{2}\right)\pi\right],
\end{equation}
where $\Phi_{0}=h/2e$ is the flux quantum. A maximum value
of $I_{c}$ is observed at $\Phi$ = 0 in a conventional SQUID, for
example, when the two terminals of the Josephson junction are placed on the
same side of the Sr$_{2}$RuO$_{4}$ crystal. In contrast, the
magnetic-field-dependent critical current shows a $\pi$ phase
shift in the opposite case, indicating odd-parity pairing
symmetry in Sr$_{2}$RuO$_{4}$. However, because of the existence
of chiral domain structures in Sr$_{2}$RuO$_{4}$, the above
experiments are hard to reproduce, and a direct observation
of chiral domains is lacking.\cite{ChiralSC_Kallin_RPP16,Sr2RuO4_MaenoY_JPCS12}
Subsequently, this $\pi$ phase shift of a similar
Josephson junction was also observed by another group.\cite{Sr2RuO4Ic_Kidw_Science06}
Chiral order parameter domains of the form $p_{x} \pm ip_{y}$ were found
in these studies, which is direct evidence for complex p-wave
pairing symmetry.\cite{Sr2RuO4Ic_Kidw_Science06}

The proposed pairing symmetry in Sr$_{2}$RuO$_{4}$ is similar to the
A-phase of superfluid $^{3}$He.\cite{Sr2RuO4_Mack_RMP03}
Half-quantum fluxoids have been predicted and should be accompanied by
MZMs in the vortices.\cite{Paired_ReadN_PRB00} As expected,
half-height magnetization steps have been successfully observed in
annular Sr$_{2}$RuO$_{4}$ samples, suggesting the existence of
half-quantum fluxoids,\cite{Halfvortex_JangJ_Science11} which can
give rise to MZMs.\cite{HalfVortexbyMZMs_Kopnin_PRB91,Paired_ReadN_PRB00} Moreover,
the differential conductance spectrum shows a broad hump, revealing
another important characteristic of supercurrents spontaneously
generated at edges.\cite{Sr2RuO4edgestate_Kash_PRL11}

The evidence of spin-triplet pairing states in Sr$_{2}$RuO$_{4}$
seems to be very strong, but open questions still remain.
First, the evidence of a superconducting gap in Sr$_{2}$RuO$_{4}$
is inconclusive, although a gap node has been evidenced by measurements of the specific heat, thermal
conductivity, and 1/T$_{1}$ as well as by the temperature dependence of
the penetration depth.\cite{Sr2RuO4_MaenoY_JPCS12,Sr2RuO4_Mack_RMP03,Sr2RuO4_Kallin_JPCM09}
A recent NMR study on Sr$_{2}$RuO$_{4}$ found a substantially suppressed spin susceptibility,
which could cause considerable controversy over its superconducting order paramenter \cite{Sr2RuO4_Pustogow_arXiv19}.
According to a theoretical work, topological
edge states may exist in nodal gap superconductors, but the edge
state is weaker than that in fully gapped systems.\cite{TopoSurfSinNodeSC_Schnyd_JPCM15}
In addition, the SOC is not very strong in Sr$_{2}$RuO$_{4}$, which causes the topological chiral
surface state to not be separated and formed by ordinary fermions
instead of chiral Majorana edge states.\cite{TSC_Sato_RPP17}
Therefore, the mixed surface states increase the difficulty of
detecting MZMs and gapless surface states, and applying a high magnetic field may be helpful for
detecting the Majorana edge states.\cite{Sr2RuO4MF_Yuji_PRL13}
More discussions on Sr$_{2}$RuO$_{4}$ can be
found in recent studies.\cite{TSC_Sato_RPP17,ChiralSC_Kallin_RPP16,Sr2RuO4_MaenoY_JPCS12}

\subsection{UPt$_{3}$}

UPt$_{3}$, a heavy fermion superconductor, is also known to be a
strong candidate for spin-triplet superconductors.\cite{TSC_Sato_RPP17,ChiralSC_Kallin_RPP16,UPt3_Joynt_RMP02}
UPt$_{3}$ has a MgCd$_{3}$-type hexagonal structure with the space group
$P6_{3}/mmc$. The large specific heat coefficient ($\gamma$) and
Fermi liquid behavior of the resistivity at low temperature illustrate
its strong electron-electron interaction as well as its heavy
fermion features.\cite{UPt3_Joynt_RMP02} These heavy fermion characteristics
can be verified by measuring the de Haas-van Alphen effect as
well as the effective masses of different bands, ranging from approximately
15 $m_{e}$ to 90 $m_{e}$. Thermodynamic measurements also
support Fermi liquid states in UPt$_{3}$. When the temperature is
below 5 K, a weak antiferromagnetic order transition emerges along
with intraplane antiferromagnetic coupling and interplane ferromagnetic coupling.

Superconductivity is observed at quite low temperatures, with
$T_{c}$ = 0.53 K, and three different superconducting phases
appear, named phase A (0.48 K $<$ T $<$ 0.53 K, low-$B$), phase B
($T$ $<$ 0.48 K, low-$B$) and phase C ($T$ $<$ 0.48 K, high-$B$).
The specific heat of UPt$_{3}$ also shows a second
transition at $T \sim$ 0.48 K, and the existence of
multicomponent phase diagrams can be verified by critical
magnetic field measurements and other experiments.\cite{UPt3_Joynt_RMP02} The internal magnetic field in
phase B has been verified by $\mu$SR measurements,\cite{UPt3uSR_Luke_PRL93} similar to the case of
Sr$_{2}$RuO$_{4}$.\cite{Sr2RuO4USR_Luke_Nature98} Moreover, the
polar Kerr effect reflects time reversal symmetry breaking in
phase B.\cite{UPt3Kerreffect_Schemm_science14} Although the
pairing symmetry is still under debate, spin triplet $f$-wave
superconductivity is believed to occur in phase B.\cite{UPt3fwave_Tsuts_JSPS13,TSC_Sato_RPP17,ChiralSC_Kallin_RPP16,UPt3Kerreffect_Schemm_science14,UPt3Twofold_Machid_PRL12,UPt3_Joynt_RMP02}
In addition, nodes in the superconducting gap are presented in all
three phases,\cite{UPt3_Joynt_RMP02} which is supposed to be
unfavorable for TSCs.\cite{TIandTSC_QiXL_RMP11} Angular-resolved
Josephson tunneling junctions experiments can probe the existence of
nodes by measuring the angular dependence of the gap, and fourfold symmetry
of the gap structure has been observed for the high temperature phase, which
is quite different from the results for the hexagonal crystal structure.\cite{UPt3GapS_Strand_Science10}
The low temperature phase seems
to exhibit different superconducting gaps, thus indicating a complex
order parameter in UPt$_{3}$.

Theoretical work has predicted the existence of two dispersive MZMs
due to the mirror chiral symmetry \cite{UPt3TSC_Tsutsumi_JSPS13}
in this nodal superconductor. This material provides another possibility
for realizing TSCs with nodal superconductors.\cite{UPt3TSCTheory_Shingo_PRB16} However, because of the
radioactive element uranium in UPt$_{3}$, performing experiments to detect the gapless surface
states and MZMs in the superconducting vortex is difficult.

\subsection{Iron-based superconductors}

Iron-based superconductors have drawn extensive attention over the past
ten years \cite{IronBasedSC} because of its high $T_{c}$, nematic
phase, orbital ordering phase, abundant antiferromagnetic phases,
etc. The rich phases may be related to the multiorbital
physics. Therefore, intrinsic hybridization of various bands
could exist in these systems and contribute to band inversions
\cite{TopologyinIronSC_HaoN_arXiv18,FeSeToP_HaoNN_PRX14} similar to
in TIs, which appeals to theorists, and nontrivial
topological states in some iron-based superconductors
have been proposed.\cite{TopologyinIronSC_HaoN_arXiv18}

Monolayer FeSe film grown on a STO substrate possesses a
very high superconducting transition temperature, with $T_{c}$
$\sim$ 77 K.\cite{FeSeSrTiO3SC_WangQY_CPL12} Intriguingly, angle-resolved
photoemission spectroscopy (ARPES) measurements revealed
1D topological edge states at the M point in an FeSe/STO sample,
where the gap forms because of band inversion caused by
SOC.\cite{TopEdgeStFeSeSTO_WangZF_NM16,FeSeToP_HaoNN_PRX14} This
interesting work provides an exciting opportunity to investigate
the coexistence of superconductivity and topological states in
iron-based superconductors. In addition, when the element Se is gradually
substituted by the element Te in monolayer
FeTe$_{1-x}$Se$_{x}$/STO, a nontrivial topological band can be
detected at the $\Gamma$ point by ARPES.\cite{FeTeSe-STOTopology_ShiX_SB17} The conduction band and
valence band at the $\Gamma$ point will touch at 33\% Se
concentration, as predicted by theoretical calculations on
single layer FeTe$_{1-x}$Se$_{x}$,\cite{FeTeSeFilm_WuXX_PRB16}
which suggests a nontrivial band topology of monolayer
FeTe$_{1-x}$Se$_{x}$/STO. In addition, Dirac-cone-type
spin-helical surface states have been observed in
FeTe$_{0.55}$Se$_{0.45}$ through high-resolution spin-resolved photoemission spectroscopy and
ARPES. Sharp ZBPs inside the vortex have also been successfully detected
and are regarded as evidence of Majorana bound states.\cite{FeTeSeMZMs_WangD_Science18}
However, in the vortex, s-wave-like quasiparticle excitations near the zero energy have
also been observed in an FeTe$_{0.55}$Se$_{0.45}$ sample.\cite{FeTe0.55Se0.45CdGM_ChenM_NC18}
Therefore, further experiments are required to verify the TSC behavior in
FeTe$_{0.55}$Se$_{0.45}$.

The progress in the topological nature of FeSe$_{1-x}$Te$_x$
inspired more investigations of other iron-based superconductors. Another FeSe-based superconductor,  Li$_{0.84}$Fe$_{0.16}$OHFeSe, was developed,\cite{LiFeOHFeSeSC_LuXF_NM15} in which the Li$_{0.84}$Fe$_{0.16}$OH layers are intercalated between FeSe layers.
Li$_{0.84}$Fe$_{0.16}$OHFeSe can be viewed as a heavily
electron-doped iron selenide superconductor, and its $T_{c}$ is
greatly enhanced from the value of 8 K in FeSe \cite{FeSe_HsuFC_PNAS08}
to $\sim$ 40 K. A full superconducting gap is obtained based on
STM and penetration depth measurements.\cite{LiFeOHFeSeFullygap_Smidman_PRB17,LiFeOHFeSeFullygapSTM_NiuXH_PRB15}
Subsequently, a ZBP was clearly observed in tunneling
spectroscopy results,\cite{LiFeOHFeSeMZM_LiuQ_PRX18} which is consistent with MZMs. The
Dirac-cone-type surface state was also confirmed by ARPES. An early
theoretical work proposed that Caroli-de Gennes-Matricon (CdGM)
bound states are generated in the vortex with separation energy
$E=\mu\Delta^{2}/F_{F}$, where $E_{F}$ is the Fermi energy, and $\mu$
= $\pm$1/2, $\pm$3/2 ... in an s-wave superconductor.\cite{CdGMS-wave_Caroli_PL64}
The discrete energy of the bound states in
a chiral p-wave superconductor will change with $\mu$ = 0, $\pm$1,
$\pm$3 ..., and the $\mu$ = 0 state is the MZM.\cite{CdGMP-wave_Volovik_TJP96} The pairing symmetry in this
system is proposed to be a bonding-antibonding s-wave state or a
nodeless d-wave state,\cite{LiFeOHFeSeZn_DuZ_NP18,GapinFeBasedSC_Hirschf_RPP11}
and Majorana edge states have not been observed thus far. More
experimental evidence is desired to confirm the topological
superconductivity.

In addition to the aforementioned superconductors, several other
iron-based superconductors have also been proposed to be promising TSCs.
Multiple topological phases have been observed in the 111 series
Li(Fe,Co)As samples by ARPES.\cite{LiFeAsTSC_ZhangP_arXiv18} CaFeAs$_{2}$ is
likewise predicted to be an ideal TSC system, whose
structure includes CaAs and FeAs layers. The CaAs layer is a
topological quantum spin Hall layer and the FeAs layer is a
superconducting layer; thus, this compound is also a promising system for
detecting MZMs.\cite{CaFeAs2TSC_WuXX_PRB15} Therefore,
iron-based superconductors may offer a potential platform to
discover TSCs,\cite{TopologyinIronSC_HaoN_arXiv18} which may open a
new direction to study iron-based superconductors.

\subsection{Cr-based superconductors}

The recently discovered Cr$_{3}$As$_{3}$-based
quasi-1D superconductors have attracted
wide attention due to their unconventional properties. Much experimental evidence for possible spin-triplet
pairing states exists.\cite{CrBasedSC_CaoGH_PM17} K$_{2}$Cr$_{3}$As$_{3}$ was
the first synthesized member, with $T_{c}$ = 6.1 K,\cite{K2Cr3As3_BaoJK_PRX15} and then,
Rb$_{2}$Cr$_{3}$As$_{3}$ ($T_{c}$ = 4.8 K)
\cite{Rb2Cr3As3_TangZT_PRB15} and Cs$_{2}$Cr$_{3}$As$_{3}$
($T_{c}$ = 2.2 K) \cite{Cs2Cr3As3_TangZT_SCM15} were discovered. Recently, Na$_2$Cr$_3$As$_3$,
with $T_c$ $\sim$ 8.6 K, was also successfully synthesized by an ion-exchange synthesis method.\cite{Na2Cr3As3_MuQG_PRM18} $T_c$ appears to decrease with
increasing radius of the $A$ ions in the $A_{2}$Cr$_{3}$As$_{3}$ series ($A$ = Na, K, Rb, Cs).

The upper critical field of K$_{2}$Cr$_{3}$As$_{3}$
can reach 37 T at 0.6 K,\cite{K2Cr3As3Hc2_Balak_PRB15,K2Cr3As3Hc2_ZuoHK_PRB17}
significantly exceeding the Pauli limit $H_{c2}$ = 1.84 $T_{c}$,
indicating an unconventional pairing state.\cite{PauliLimit_Clogston_PRL62}
Moreover, Cr-spin ferromagnetic fluctuations, as measured by NMR, emerge above
$T_{c}$,\cite{Rb2Cr3As3NMR_YangJ_PRL15,K2Cr3As3NMR_ZhiHZ_PRL15}
and a $\mu$SR study showed that a weak spontaneous internal magnetic
field is detected below $T_{c}$,\cite{K2Cr3As3USR_Adroja_PRB15} both of which
suggest potential spin-triplet pairing states in these
Cr-based superconductors. Other experiments, such as penetration depth measurements, suggest
gap nodes in the superconducting gap function.\cite{K2Cr3As3node_PangGM_PRB,CrBasedSC_CaoGH_PM17}
Because of the air-sensitive feature of the Cr-based samples, performing
further measurements to investigate their topological nature is very difficult.

\section{Pressure induced superconductivity}

The application of high pressure, as a clean and effective method to tune the lattice
and electronic structures without inducing disorder, is usually employed to investigate
phase transitions and synthesize certain compounds.\cite{DAChighpressure_JA_RMP83}
Especially in the study of superconductors, employing high pressure has been proven
to be useful.\cite{highpressure_MaoHK_RMP18} The highest recorded $T_{c}$ was realized
in hydrogen-containing materials by applying high pressure.\cite{H2SSC_Drozd_Nature16,HPtoroomtSC_Gor_RMP18}
Therefore, high pressure has also been widely used to explore
superconductivity in topological materials.\cite{DAChighpressure_JA_RMP83} Although superconductivity has
been induced by high pressure in all types of topological materials, including
topological (crystalline) insulators, DSMs, and WSMs, much work remains before either TSCs
or MZMs can be observed with this method. Probing pairing states and gapless
surface states under high pressure condition is difficult. However, exploration of
topological superconductivity in topological materials
using the high pressure technique is worthy of continuous endeavors.

\subsection{Topological insulators under pressure}

TSCs are fully gapped in the bulk state, similar to TIs, but have a Majorana
bound state instead of the Dirac surface state in
TIs.\cite{TIandTSC_QiXL_RMP11} Many theoretical predictions and
experiments have been performed to investigate 2D TIs, such as HgTe/CdTe quantum well structures,\cite{HgCdTe2DTI_RothA_science09} or 2D Dirac surface states in
three-dimensional (3D) TIs, such as Bi$_{1-x}$Sb$_{x}$,\cite{Sb1-xBixTI_Hsieh_nat08} Bi$_{2}$Te$_{3}$,\cite{Bi2Te3TI_ChenYL_science09} Bi$_{2}$Se$_{3}$,\cite{Bi2Se3_XiaY_NatP09}
Sb$_{2}$Te$_{3}$ \cite{Sb2Te3_HsiehD_PRL09} and others.\cite{TI_Hasan_RMP10}
Based on the studies of TIs, tuning TIs such that they exhibit superconductivity by
applying high pressure has become an attractive option and has been
performed by many groups to discover possible TSCs.

Bi$_{2}$Te$_{3}$ has been identified as a TI based on ARPES \cite{Bi2Te3TI_ChenYL_science09}
and displays a superconducting transition of approximately 3 K under 4 GPa.\cite{Bi2Te3SC_ZhangJL_PNAS11,Bi2Te3Hall_ZhangC_PRB11} $T_{c}$ increases at approximately 8 GPa,
accompanied by a structural transition from the ambient-pressure phase
(a rhombohedral $R$-$3m$ structure) to the high-pressure phase (a monoclinic $C2/m$ structure).\cite{Bi2Te3_MK_PRB14,Bi2Te3SC_ZhangJL_PNAS11,Bi2Te3Hall_ZhangC_PRB11}
According to density functional theory (DFT) calculations, the topologically nontrivial
band structure and surface states still exist at 4 GPa,\cite{Bi2Te3SC_ZhangJL_PNAS11}
which may induce Majorana fermions in the surface states due to the proximity effect
of the bulk superconducting states.\cite{SCpeonTI_FuL_PRL08}

\begin{figure}[!thb]
\begin{center}
\includegraphics[width=3.5in]{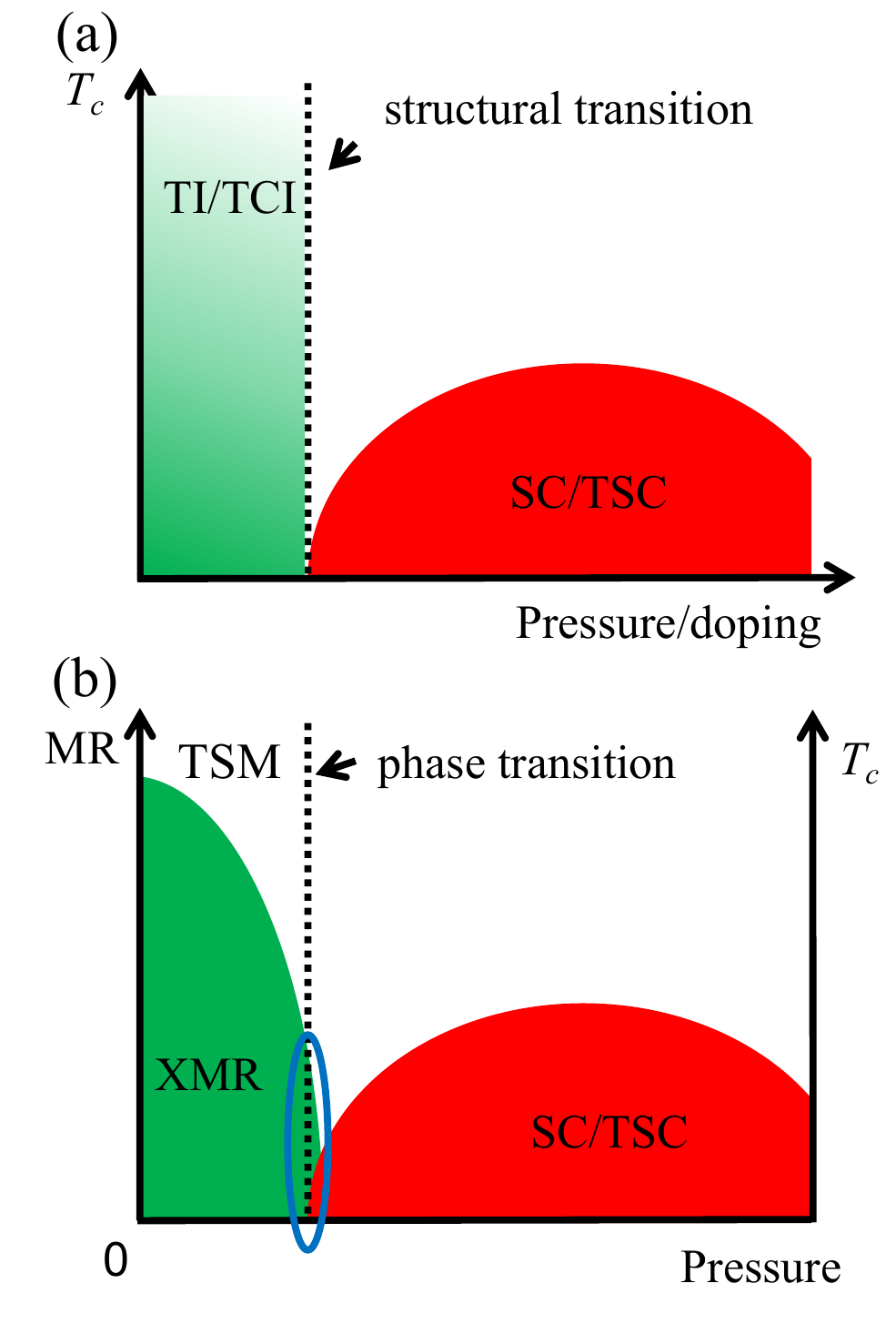}
\end{center}
\caption{\label{TItoSC} Sketched phase diagrams of TIs/TCIs (a) and TSMs (b) under
pressure or doping. SC, TCI, and MR represent superconductivity,
topological crystalline insulator and magnetoresistance, respectively.
(a) A superconducting state (red region) is usually induced by pressure
after a structural transition. No structural transition emerges in the TCI/TI doping cases.
For TSMs, whether the suppression of XMR or a structural transition drives
the system into a superconducting state is still an open issue.
XMR and SC may coexist in the region marked by the ellipse.   }
\end{figure}

\begin{table*}
\tabcolsep 0pt \caption{\label{HPSC} Summary of high-pressure-induced superconductivity in topological materials.
Superconductivity usually appears along with a structural
transition under pressure. In the $T_{c}$ column, the pressure
under which superconductivity begins to emerge is listed in brackets.
In the Phase II and Phase III columns, the pressure at which the
structural transition begins is also listed in brackets.
(In this table, TI, DSM, WSM, TSM and TPSM represent topological
insulator, Dirac semimetal, Weyl semimetal, topological semimetal
and triply degenerate nodal point semimetal, respectively.
Pressure and temperature are simplified as P and T,
respectively.) } \vspace*{-12pt}
\begin{center}
\def\temptablewidth{2.05\columnwidth}
{\rule{\temptablewidth}{1pt}}
\begin{tabular*}{\temptablewidth}{@{\extracolsep{\fill}}cccccc}
Material              &Type   &$T_{c}$   &Phase I (Ambient P)   &Phase II $\quad$  &Phase III  \\ \hline
Bi$_{2}$Se$_{3}$ \citep{Bi2Se3SC_Kevin_PRL13}  & TI &0.3-7 K (11.9 GPa)   &$R$-$3m$  &$C2/m$ (10 GPa)   &bcc-like C2/m (28 GPa)\\
Bi$_{2}$Te$_{3}$ \cite{Bi2Te3SC_ZhangJL_PNAS11,Bi2Te3Hall_ZhangC_PRB11,Bi2Te3_MK_PRB14} &TI &2.6-9.5 K ($\sim$3 GPa) &$R$-$3m$ &$C2/m$ ($\sim$8 GPa) &--\\
Sb$_{2}$Te$_{3}$ \cite{Sb2Te3_ZhuJ_ScientificR,Sb2Te3P_ZhaoJG_IC11} &TI &3-7.3 K(4 GPa) &$R$-$3m$ &$C2/m$ (9.3 GPa) &$Im$-$3m$ (19.8 GPa)\\
BiTeI \cite{BiTeISC_QiYP_AM17} &TI &$\sim$1-5.2 K ($\sim$10 GPa)  &$P3m1$ &$Pnma$ (8.8 GPa) &$P4/nmm$ (18.9 GPa) \\
BiTeBr \cite{BiTeISC_QiYP_AM17} &TI &$\sim$1-4.8 K ($\sim$14 GPa)  &$P3m1$ &$Pnma$ (6.6 GPa) &$P4/nmm$ (19.1 GPa) \\
BiTeCl \cite{BiTeClSC_YangJJ_PRB16} &TI &$\sim$4-8 K ($\sim$11 GPa) &$P6_{3}mc$ &possible $Pnma$ (5 GPa) &possible cubic (39 GPa) \\
Cd$_{3}$As$_{2}$ \cite{Cd3As2_HeLP2016QuantMater} &DSM &2-4 K (8.5 GPa) &$I4_{1}/acd$  &$P2_{1}/c$ (3.5 GPa) &--  \\
ZrTe$_{5}$ \cite{ZrTe5HP_zhouYH_PANS} &DSM &1.8-6 K (6.7 GPa)  &$Cmcm$  &$C2/m$ (6 GPa)  &$P$-$1$ (21.2 GPa)\\
HfTe$_{5}$ \cite{HfTe5SC_QiY_PRB16} &DSM &1.8-4.8 K (5 GPa)  &$Cmcm$ &$C2/m$ (2.2 GPa)  &$P$-$1$ (14.8 GPa) \\
TaP \cite{TaPSC_LiY_npjQM17} &WSM &1.8-3 K (70 GPa) &$I4_{1}md$ &$P$-$6m2$ (70 GPa) &-- \\
MoTe$_{2}$ \cite{MoTe2_Qiy_NC16} &WSM &0.1-8.2 K (Ambient P) &$P2_{1}/m$ (High T) &$Pmn2_{1}$ (Ambient P, low T)  &-- \\
LaBi \cite{LaBiSC_TaftiFF_PRB17} &TSM &4-8 K (3.5 GPa) &$Fm$-$3m$ &$P4/mmm$ (11 GPa) &-- \\
WTe$_{2}$ \cite{WTe2_DFKang_NatC,WTe2_XCPan_NatC} &WSM &$\sim$3-7 K (2.5 GPa) &$Pmn2_{1}$ &-- &-- \\
MoP \cite{MoPHPSC_ChiZ_npj18} &TPSM   &2.5-4 K (30 GPa) &$P$-$m62$  &-- &-- \\
NbAs$_{2}$ \cite{NbAs2SC_Liyp} &TSM   &2-2.6 K (12.8 GPa) &$C2/m$  &-- &-- \\
\end{tabular*}
{\rule{\temptablewidth}{1pt}}
\end{center}
\vspace*{-18pt}
\end{table*}

A more intriguing TI compound, Bi$_{2}$Se$_{3}$, could become
superconducting not only through charge doping \cite{CuxBi2Se3_HorYS_PRL10,SrxBi2Se3_Shruti_PRB15}
but also through high pressure application.\cite{Bi2Se3SC_Kevin_PRL13}
When pressure is applied up to 11 GPa, a superconducting transition is successfully
observed, accompanied by a structural transition from the $R$-$3m$ phase
to the $C2/m$ phase, similar to in Bi$_{2}$Te$_{3}$.\cite{Bi2Te3SC_ZhangJL_PNAS11}
$T_{c}$ can reach a maximum of 7 K below 50 GPa when a body-centered cubic structure,
similar to the $C2/m$ structure, appears at 28 GPa.\cite{Bi2Te3_MK_PRB14,Bi2Se3struc_Vilaplana_PRB11}
The quasi-linear temperature-dependent $H_{c2}$ exceeds both
the orbital and Pauli limits, which indicates an unconventional
pairing state in Bi$_{2}$Se$_{3}$ under pressure.\cite{Bi2Se3SC_Kevin_PRL13}
Sb$_{2}$Te$_{3}$ is another member of
this TI family, and it becomes superconducting at 3 K when the applied
pressure is increased up to 4 GPa.\cite{Sb2Te3_ZhuJ_ScientificR} The $C2/m$ structure seems to be more stable above 12.9 GPa in Sb$_{2}$Te$_{3}$.\cite{Sb2Te3P_ZhaoJG_IC11} According to
DFT calculations using an experimental lattice at high pressure,
the Dirac surface states remain at 6.9 GPa in
coexistence with superconductivity, which strongly suggests that
this material is a promising TSC candidate.

Another TI material family, BiTeX (X = Cl, Br, and I), consists of
inversion-asymmetric TIs, which are usually
called Rashba semiconductors because of the giant Rashba-type spin
splitting.\cite{BiTeIAPRES_Ishizaka_natM11,BiTeIAPRES_Sakano_PRL13}
Robust Dirac surface states and large Rashba splitting have been revealed by
ARPES.\cite{BiTeCl_ChenYL_NatP13,BiTeIAPRES_Sakano_PRL13}
A maximum $T_{c}$ of 5.2 K is reached at 23.5 GPa in BiTeI after two
structural transitions.\cite{BiTeISC_QiYP_AM17} Interestingly, a nontrivial topology
of the band structure is also obtained in the high pressure phase according to
DFT calculations. In addition, similar pressure-induced
superconducting phases have also been detected in BiTeBr
\cite{BiTeISC_QiYP_AM17} and BiTeCl.\cite{BiTeClSC_YangJJ_PRB16}
Thus, members of the BiTeX family may be promising TSCs under pressure.

The properties of TIs under pressure are summarized in Table.\ref{HPSC}.
Fig.\ref{TItoSC}(a) illustrates a sketched phase diagram of TIs under pressure,
characterized by a structural transition before entering the superconducting phase.
Only in some cases does the superconducting phase appear before the structural transition, for
example, in Bi$_{2}$Te$_{3}$ \cite{Bi2Te3SC_ZhangJL_PNAS11} and
Sb$_{2}$Te$_{3}$.\cite{Sb2Te3_ZhuJ_ScientificR} Therefore, determining whether
there exist nontrivial topological states in the superconducting state requires further
measurements, but the high pressure conditions make investigating topological superconductivity difficult.

\subsection{Topological semimetals under pressure}

The successful discovery of TIs has stimulated tremendous research on 3D DSMs with a 3D linear energy dispersion relation. Dirac nodes with fourfold degenerate crossings in the Brillouin zone have been confirmed in Na$_{3}$Bi,\cite{Na3Bi_LiuZK_science14} Cd$_{3}$As$_{2}$,\cite{Cd3As2_LiuZK_NatM14} ZrTe$_{5}$,\cite{ZrTe5ARPES_LiQ_NP16} etc. Intriguingly, a Dirac node will be split into two two-fold degenerate crossings with opposite chirality when a magnetic field is applied. These band crossings are named Weyl nodes \cite{WSMWanXG_PRB} and exist in WSMs, where time reversal or inversion symmetries are broken because of magnetism or the noncentrosymmetric structure.\cite{WeylSemi_WHM_PRX} The Fermi arc formed by quasiparticle excitations of Weyl fermions was first confirmed in noncentrosymmetric transition-metal monophosphides (TaAs,\cite{TaAs_Hasan_Science} TaP,\cite{TaP_XuSY_SciAd15} NbAs,\cite{NbAs_XuSY_NatP15} NbP \cite{NbPFA_XDF_CPL15}). The negative magnetoresistance caused by chiral Weyl nodes when the electric field is parallel to the magnetic field is regarded as a very important feature of WSMs.\cite{WeylSemi_WHM_PRX,NbP_WZ_PRB} Several other effects, such as the current jetting effect, \cite{currentjeting_HuJ_PRL05} could also cause a similar negative magnetoresistance.\cite{NMRinWSMs_dRD_NJP16,NMR_YPL_FP17}

Similar to TIs, DSMs and WSMs are also good platforms for exploring
TSCs, and superconductivity can be induced by tuning the carrier
density using pressure or doping. As expected, pressure-induced superconductivity
is detected at 2 K in Cd$_{3}$As$_{2}$, a typical DSM, under 8.5 GPa.\cite{Cd3As2_HeLP2016QuantMater}
The superconductivity is quite robust and survives up to 50.9 GPa, with a maximal $T_{c}$ of 4.0 K
under 21.3 GPa. Before the emergence of superconductivity in
Cd$_{3}$As$_{2}$, the tetragonal $I4_{1}/acd$ structure changes to
a monoclinic $P2_{1}/c$ phase at approximately 3.5 GPa.
Superconductivity was observed at approximately 6 GPa in another DSM,
ZrTe$_{5}$, where a structural transition from the ambient
$Cmcm$ phase to the monoclinic $C2/m$ phase occurs.\cite{ZrTe5HP_zhouYH_PANS}
When the triclinic $P$-$1$ phase becomes stable
at higher pressures above 21.2 GPa, a maximum $T_{c}$ of $\sim$ 6.0 K
is obtained under 30 GPa. However, this structural transition, as
supported by high-pressure X-ray diffraction measurements, may lead to the breakdown
of the 3D DSM state in Cd$_{3}$As$_{2}$ \cite{Cd3As2break_ZhangS_PRB15}
and ZrTe$_{5}$.\cite{ZrTe5HP_zhouYH_PANS} The effect of such a
structural transition on topological properties is still
ambiguous. Some theoretical works suggest that this symmetry
breaking during the structural transition may be favorable for the formation of
a topological superconducting phase with surface Majorana fermions in
Cd$_{3}$As$_{2}$.\cite{Cd3As2TSCTheory_Kob_PRL15} HfTe$_{5}$ is
an compound analogous to ZrTe$_{5}$ in which two structural
transitions occur, with one superconducting phase at 5 GPa.\cite{HfTe5SC_QiY_PRB16}
First-principles calculations of these two high pressure phases showed
the trivial properties in HfTe$_{5}$ under pressure.

The search for TSCs and Majorana fermions in WSMs is also
underway \cite{WeylSCMF_SauJ_PRB12} through application of high pressure
or the proximity effect between an s-wave superconductor and the
surface states of a WSM.\cite{proximityEfWeylSC_ChenA_PRB16}
When NbAs is subjected to pressure, no superconductivity exists up to 26 GPa while the
ambient-pressure structure is retained.\cite{NbAs_ZhangJ_CPL15}
However, if a higher pressure is applied to TaP, a superconducting transition
is observed above 70 GPa, accompanied by a structural transition from $I4_{1}md$ to
$P$-$6m2$.\cite{TaPSC_LiY_npjQM17} Interestingly,
superconductivity becomes more robust and even remains at 2 GPa in
pressurized TaP during the pressure decreasing process.
DFT calculations indicate that a new WSM state appears in the $P$-$6m2$ phase,
different from the ambient phase. This concurrence of superconductivity and the WSM state
may result in a TSC for this WSM.

Type-II WSMs, such as WTe$_{2}$,\cite{WTe2WSM_WuL_PRB16} MoTe$_{2}$,\cite{MoTe2WSM_Dengk_NP16} Mo$_{x}$W$_{1-x}$Te$_{2}$,\cite{MoxWTe2SM_ChangTR_nc16} and
Ta$_{3}$S$_{2}$,\cite{Ta3S2TpeIIWS_ChangG_SciAdv16} have drawn increasing attention due to the tilted and
protected Weyl-type band crossings.\cite{typeIIWeylS_SoluAA_nat15}
Exploring TSCs among type-II WSMs is also attractive because they
are more common than type-I WSMs. In the type-II WSM WTe$_{2}$, an extremely large magnetoresistance (XMR)
was reported due to its electron-hole compensation.\cite{WTe2_Ali_Nat14} With increasing pressure, superconductivity
is induced by suppression of XMR in WTe$_{2}$,\cite{WTe2_DFKang_NatC,WTe2_XCPan_NatC} as shown in
Fig.\ref{TItoSC}(b). The sign of the Hall coefficients changes without
a structural transition when superconductivity emerges in the
vicinity of 10.5 GPa. Thus, reasonably, the mismatch between the electron and hole
carrier densities at high pressure could contribute to the
suppression of XMR.\cite{WTe2_DFKang_NatC} The explanation of the occurrence of
superconductivity in WTe$_{2}$ under pressure is related to the
increase of the density of states.\cite{WTe2_XCPan_NatC} Another
type-II WSM, MoTe$_{2}$, shows a superconducting transition at 0.1
K even under ambient pressure.\cite{MoTe2_Qiy_NC16} After a
structural transition at 240 K from the T$'$ phase to the T$_{d}$ phase,
WSM states could be observed, as expected, in the T$_{d}$ phase by ARPES.\cite{MoTe2ARPES_DengK_NP16,TdMoTe2ARPES_HuangL_NM16} In addition, a high pressure will increase $T_{c}$ to a maximum value
of 8.2 K, and the structural transition temperature will decrease
with increasing pressure.\cite{MoTe2_Qiy_NC16}

Moreover, superconductivity under pressure has also been detected
in several other topological semimetals, such as LaBi,\cite{LaBiARPES_NayakJ_ncom17,LaBiSC_TaftiFF_PRB17} MoP,\cite{triply_BQLv_nature17,MoPHPSC_ChiZ_npj18} and NbAs$_{2}$.\cite{NbAs2SC_Liyp,NbAs2DiracNL_YMShao_PNAS19}
The high-pressure-induced superconductivity in these topological materials is
seemingly accompanied by crystal structure symmetry breaking, as seen in
Fig.\ref{TItoSC}(b) and Table.\ref{HPSC}, which usually changes the
topological features. However, superconductivity and nontrivial
topological properties should simultaneously exist in a
potential TSC. High pressure conditions make performing
crucial experiments, such as ARPES, to test the
nontrivial topological properties difficult, and pressure-induced structural transitions
make such experiments even more complicated. Consequently, investigating the topological semimetals that become superconducting without a structural transition under pressure, such as
WTe$_{2}$,\cite{WTe2_DFKang_NatC,WTe2_XCPan_NatC} MoP,\cite{triply_BQLv_nature17,MoPHPSC_ChiZ_npj18} and  NbAs$_{2}$,\cite{NbAs2DiracNL_YMShao_PNAS19,NbAs2SC_Liyp} is relatively more meaningful.
For example, pressure-induced superconductivity is obtained at 2.63 K and
12.8 GPa in the topological semimetal NbAs$_{2}$ without a
structural transition.\cite{NbAs2SC_Liyp} A linear Dirac dispersion was recently detected
in NbAs$_{2}$ through magneto-optical measurements.\cite{NbAs2DiracNL_YMShao_PNAS19}
A negative magnetoresistance is widely observed in NbAs$_{2}$,
\cite{NbAs2family_YupengLi_arXiv,NbAs2_ShenB_PRB} as predicted.\cite{NbAs2Weyl_GreschD_NJP17}
In addition, as a semimetal with triply degenerate nodal points,\cite{triply_BQLv_nature17} MoP
exhibits superconductivity when subjected to pressure above 30 GPa,
while no structural transition occurs up to 60 GPa.\cite{MoPHPSC_ChiZ_npj18}
DFT calculations with SOC uncovered the coexistence of Weyl points and triply
degenerate nodal points under high pressure. In these systems,
topological properties and superconductivity coexist under pressure,
which may be a new platform to explore TSC candidates.

\section{Tip-induced superconductors}

Tip-induced superconductivity in topological
semimetals has been recently observed in hard point contact experiments.\cite{TipSC_WangH_SB18}
Usually, hard point contact measurement is a useful method to study the gap structure of
superconductors. This technique has become a new method for tuning
nonsuperconducting materials into superconducting states, similar to
other modulation methods, such as electric field gating, high
pressure application, and chemical doping. Obviously, this technique has the advantage
that the point contact spectroscopy (PCS) can be simultaneously
measured when the sample is tuned. Thus, hard point contact
measurement of topological semimetals offers a new way to trigger
and detect topological superconductivity.

When the hard tip method is implemented on the DSM
Cd$_{3}$As$_{2}$, the resistance begins to drop at 3.9 K,
suggesting a superconducting transition.\cite{Cd3As2tipSC_WangH_NM16}
Another group reported a similar experiment in which Cd$_{3}$As$_{2}$ shows
$T_{c}$ = 5.8 K.\cite{Cd3As2tipSC_Aggar_NM16} This difference in $T_c$ may
originate from the pressure applied by the hard tip,
and the actual pressure on the sample is difficult to measure in such
experiments. Charge transfer may also exist between the tip
and the sample, which would lead to the charge doping effect,
similar to the chemical doping case. When a magnetic field is employed
perpendicular to the plane on which the pressure is applied,
the magnetic-field-dependent $T_{c}$ shows a concave curvature
in the low magnetic field regime, indicating unconventional
superconductivity.\cite{Cd3As2tipSC_Aggar_NM16} In the PCS results of
the above two reports, superconducting gaps are successfully observed with ZBPs at zero
energy. The ZBP behaviors at various temperatures and magnetic
fields suggest the existence of MZMs in Cd$_{3}$As$_{2}$.
A tip-induced superconducting transition, with $T_{c}$ = 5.9 K, and ZBPs have also
been observed in the WSM TaAs.\cite{TipSC_WangH_SB18} However, further
theoretical and experimental studies are highly desired to
understand the exact mechanisms that occur in the tip contact region
and how to precisely analyze PCS results.

\section{Doped topological materials}

Chemical doping, including doping and intercalation, is
widely used to induce superconductivity in electron correlation
systems and to study interesting physics-like phase transitions
by modulating the carrier density density/Fermi level or
introducing chemical pressure. For example, superconductivity was realized
in fluorine-doped LaOFeAs through electron doping, thus beginning the era of iron-based superconductors.\cite{LaOFeAsF_Kamihara_JACS08} Chemical pressure can also be
induced by doping with elements with smaller radii, such as P-for-As
doping in BaFe$_{2}$As$_{2}$.\cite{PdopingSC_JiangS_JPCM09} In
addition, some fantastic physical phenomena are observed in
magnetic-element-doped systems, such as the quantum anomalous Hall
effect in a chromium-doped TI, where the time reversal symmetry is
broken,\cite{QAHinCrxBi2Se3_ChangCZ_Science13} and dilute
ferromagnetism in dilute magnetic semiconductors and dilute
magnetic oxides.\cite{DMS_DietlT_NatM10} Therefore, doping
or intercalation of a TI is naturally an interesting approach to
exploring potential TSCs.\cite{SCdopedTM_Sasaki_PhysC15}

\begin{table}
\tabcolsep 0pt \caption{\label{dopingSC} Summary of doping-induced
superconductivity in topological insulators (TIs) or topological
crystalline insulators (TCIs). $T_{c}^{max}$ is the maximum
value of the superconducting transition temperature for various
doping contents $x$.  } \vspace*{-12pt}
\begin{center}
\def\temptablewidth{1.0\columnwidth}
{\rule{\temptablewidth}{1pt}}
\begin{tabular*}{\temptablewidth}{@{\extracolsep{\fill}}cccc}
Material                    &Type  &$T_{c}^{max}$  &$x$    \\ \hline
Cu$_{x}$Bi$_{2}$Se$_{3}$ \cite{CuxBi2Se3_HorYS_PRL10,CuxBi2Se3ECQ_Kriener_PRB11}    &TI     &3.8 K        &0.09 $<x<$ 0.64        \\
Sr$_{x}$Bi$_{2}$Se$_{3}$ \cite{SrxBi2Se3SC_Liuzh_JACS15,SrxBi2Se3_Shruti_PRB15}    &TI     &2.9 K        &0.058 $<x<$ 0.1       \\
Nb$_{x}$Bi$_{2}$Se$_{3}$ \cite{Nb0.25Bi2Se3_QiuYS_arXiv15}    &TI     &3.2 K        &0.25    \\
Tl$_{x}$Bi$_{2}$Se$_{3}$ \cite{Tl0.6Bi2Te3SC_WangZW_CM16,Tl0.5Bi2Te3ARPES_Trang_prb16}    &TI     &2.28  K        &0.6    \\
Cu$_{x}$(PbSe)$_{5}$(Bi$_{2}$Se$_{3}$)$_{6}$ \cite{Cux(PbSe)5(Bi2Se3)6_Sasaki_PRB14} &TI     &2.85 K        &0.3 $<x<$ 2.3     \\
Sn$_{1-x}$In$_{x}$Te \cite{Sn1-xInxTeSC_Ehhanced_PRB09,Sn0.6In0.4Te_Balak_PRB13}    &TCI     &4.7 K        &0.017 $<x<$ 0.4        \\
(Pb$_{0.5}$Sn$_{0.5}$)$_{1-x}$In$_{x}$Te \cite{(PbSn)1-xInxTeSC_ZhangRD_PRB14} &TCI     &4.7 K &0.1 $<x<$ 0.3      \\
\end{tabular*}
{\rule{\temptablewidth}{1pt}}
\vspace*{-18pt}
\end{center}
\end{table}

\subsection{Doping of topological insulators}

Bi$_{2}$Se$_{3}$ is one of the most studied TIs.\cite{Bi2Se3_XiaY_NatP09} Copper elements can be intercalated in
the van der Waals gaps between Bi$_{2}$Se$_{3}$ layers to obtain
superconductivity, with $T_{c}$ = 3.8 K, in
Cu$_{x}$Bi$_{2}$Se$_{3}$, and the content of intercalated Cu is
0.12 $\leq x \leq$ 0.15 in the samples prepared by the solid phase
reaction.\cite{CuxBi2Se3_HorYS_PRL10} The content of Cu in
Cu$_{x}$Bi$_{2}$Se$_{3}$ synthesized by the electrochemical
intercalation technique can be better controlled, and the superconducting shielding
fractions are as large as $\sim50\%$.\cite{CuxBi2Se3ECQ_Kriener_PRB11} These interesting structures and
physics immediately drew extensive attention. A full energy gap
in the bulk superconductivity was observed by temperature-dependent specific heat
\cite{CuxBi2Se3fullgap_Kriener_PRL11} and STM measurements.\cite{CuxBi2Se3sWave_LevyN_PRL13}
More intriguingly, ARPES measurements revealed the existence of
nontrivial surface states.\cite{CuxBi2Se3SS_LahoudE_PRB13} The
upper critical field data measured under various
pressures are not consistent with the behavior of a weak coupling, orbital-limited,
spin-singlet superconductor, suggesting a spin-triplet or an anisotropic
spin-singlet state in Cu$_{0.3}$Bi$_{2}$Se$_{3}$.\cite{CuxBi2Se3HPSC_BayTV_PRL12}
Moreover, an evident ZBP caused by Andreev bound states consisting of Majorana fermions was
obtained with the `soft' point contact technique.\cite{CuxBi2Se3ZBCP_Sasaki_PRL11} All of these experiments
indicate that Cu$_{x}$Bi$_{2}$Se$_{3}$ is a good candidate for
time-reversal-invariant TSCs with massless Majorana fermions in the
surface states.\cite{CuxBi2Se3Theory_FuL_PRL10}

More interestingly, nematic superconductivity has been observed in Cu$_{x}$Bi$_{2}$Se$_{3}$,
where the spin rotation symmetry is broken below the superconducting transition temperature.\cite{CuxBi2Se3NMR_MatanoK_NP16,CuxBi2Se3Thermo_YonezawaS_NP17}
When an external magnetic field is applied to the hexagonal plane, two-fold symmetry
of the Knight shift is detected by $^{77}$Se NMR measurements below $T_{c}$ = 3.4 K
in Cu$_{0.3}$Bi$_{2}$Se$_{3}$, which possesses a three-fold rotational symmetry
lattice.\cite{CuxBi2Se3NMR_MatanoK_NP16} This symmetry breaking behavior suggests
a pseudo-spin-triplet state of Cooper pairs. Further measurements, such as specific heat,
upper critical field \cite{CuxBi2Se3Thermo_YonezawaS_NP17} and STM measurements,\cite{CuxBi2Se3_TaoR_PRX18}
confirmed this spontaneous rotational symmetry breaking (RSB) superconductivity,
which was named as a new class of nematic superconductivity with odd-parity pairing.\cite{CuxBi2Se3Thermo_YonezawaS_NP17,CuxBi2Se3_FuL_PRB} Nematic superconductors
in which spontaneous RSB exists in the amplitude factor of the superconducting gap are
completely different from superconductors hosting spontaneous RSB in the phase
factor of the superconducting gap, which can be observed by phase-sensitive junction techniques.\cite{CuprateSC_Tsuei_RMP00}

Bulk superconductivity with a large superconducting volume fraction
($\sim91.5\%$) has also been detected in Sr$_{x}$Bi$_{2}$Se$_{3}$, which exhibits
a maximum $T_{c}$ = 2.5 K when $x$ $\sim$ 0.06,\cite{SrxBi2Se3SC_Liuzh_JACS15}
and a higher $T_{c}$ of 2.9 K was found in an $x$ = 0.1 sample.\cite{SrxBi2Se3_Shruti_PRB15}
Instead of Cu and Sr intercalation, niobium intercalation in Bi$_{2}$Se$_{3}$ can
enhance $T_{c}$ to 3.2 K, with a 100\% superconducting volume fraction.\cite{Nb0.25Bi2Se3_QiuYS_arXiv15}
The intercalation of Cu in the interesting layer-structured TI
(PbSe)$_{5}$(Bi$_{2}$Se$_{3}$)$_{6}$, which is viewed as a
heterostructure with PbSe units and Bi$_{2}$Se$_{3}$ units,\cite{(PbSe)5(Bi2Se3)6_Naka_PRL12}
results in superconductivity with a maximum $T_{c}$ = 2.85 K.\cite{Cux(PbSe)5(Bi2Se3)6_Sasaki_PRB14}
Unconventional superconductivity with a nodal gap in
Cu$_{x}$(PbSe)$_{5}$(Bi$_{2}$Se$_{3}$)$_{6}$ was verified by
specific heat \cite{Cux(PbSe)5(Bi2Se3)6_Sasaki_PRB14} and ARPES
measurements,\cite{Cux(PbSe)5(Bi2Se3)6ARPES_Nakay_PRB15} and the
topological state and superconductivity with a nodal gap may be able to
coexist.\cite{TSC_Sato_RPP17} Similar to Cu$_{x}$Bi$_{2}$Se$_{3}$,
nematic order with two-fold symmetry was also revealed in the bulk states of Sr$_{x}$Bi$_{2}$Se$_{3}$,\cite{SrxBi2Se3NSC_PanY_SR16,SrxBi2Se3NSC_DuG_SCP17}
Nb$_{x}$Bi$_{2}$Se$_{3}$ \cite{NbxBi2Se3NSC_AsabaT_PRX17,Nb0.25Bi2Se3_ShenJ_npjQM17}
and Cu$_{x}$(PbSe)$_{5}$(Bi$_{2}$Se$_{3}$)$_{6}$.\cite{Cux(PbSe)5(Bi2Se3)6Nematic_Andersen_PRB18}
Therefore, these intercalated systems offer a good platform to study
this new type of TSC, and various theories also support this
point of view.\cite{CuxBi2Se3Theory_FuL_PRL10,CuxBi2Se3_FuL_PRL,turning_wanxg_NC}

\subsection{Doping of topological crystalline insulators}

Unlike TIs, TCIs exhibit topological properties that originate from crystal
symmetries when SOC is not taken into consideration.\cite{TCI_Ful_PRL11}
For example, gapless surface states in the TCI SnTe with the $Fm$-$3m$ space group
are protected by the mirror symmetry of the crystal with respect to the \{110\}
mirror plane. A gap in the surface states will open if the mirror symmetry is
broken by an in-plane magnetic field.\cite{SnTeTheory_Hsieh_NC12}
Subsequently, ARPES measurements revealed a metallic Dirac-cone
surface state in SnTe,\cite{SnTeARPES_Tanaka_NP12} which was successfully confirmed as a
TCI. When indium is doped into SnTe, superconductivity is observed
in Sn$_{1-x}$In$_{x}$Te for $x >$ 1.7\%, while the ferroelectric
structural phase transition (at 95 K in SnTe) is completely
suppressed for $x >$ 4\%.\cite{Sn1-xInxTeSC_Ehhanced_PRB09}
$T_{c}$ is enhanced to 4.7 K at $x$ = 0.4, and the
Ginzburg-Landau parameter $\kappa$ = 56.4(8) estimated from the
upper and lower critical fields suggests a type-II superconductor.\cite{Sn0.6In0.4Te_Balak_PRB13}
A full superconducting gap in Sn$_{1-x}$In$_{x}$Te has been established by thermal conductivity,\cite{Sn1-xInxTethermocond_HeLP_PRB13} specific heat
\cite{Sn1-xInxTe_Novak_PRB13} and $\mu$SR measurements.\cite{Sn1-xInxTeUSR_Saghir_PRB14}
Intriguingly, PCS experiments on a $x = 0.045$ sample uncovered a large
magnitude zero-bias conductance peak, which was caused by the
surface Andreev bound states.\cite{Sn1-xInxTeTSC_Sasaki_PRL12}
High-resolution ARPES measurements indicate that the topological surface
state is retained in Sn$_{1-x}$In$_{x}$Te.\cite{Sn1-xInxTeARPED_Sato_PRL13}
All of these data provide evidence for topological superconductivity in Sn$_{1-x}$In$_{x}$Te.

Because of the hole-doped nature of SnTe, Pb doping can tune
SnTe or SnSe to either n-type or p-type conductivity, and a large
doping range may induce topological phase transitions.\cite{Pn1-xSnxTetheory_Tanaka_PRB13,SnTeTCI_Tanaka_NP12,SnTeTheory_Hsieh_NC12,Pn1-xSnxSeTheoryArpes_Dziawa_NM12}
A topological phase transition indeed occurs in Pb$_{1-x}$Sn$_{x}$Te
at $x$ = 0.25 from a trivial band insulator phase (PbTe) to a TCI phase (SnTe).\cite{Pn1-xSnxTetheory_Tanaka_PRB13,SnTeTCI_Tanaka_NP12,SnTeTheory_Hsieh_NC12}
The analogous compound Pb$_{1-x}$Sn$_{x}$Se also possesses a TCI state at
$x$ =0.23,\cite{Pn1-xSnxSeTheoryArpes_Dziawa_NM12} and further transport experiments
revealed its massive Dirac bulk states.\cite{Pn1-xSnxSetrans_LiangT_NC13}
Moreover, interestingly, the indium substituted compound
(Pb$_{0.5}$Sn$_{0.5}$)$_{1-x}$In$_{x}$Te displays
superconductivity in the range of 0.1 $< x < $0.3, with maximum
$T_{c}$ = 4.7 K.\cite{(PbSn)1-xInxTeSC_ZhangRD_PRB14}
The superconducting state is fully gapped without any in-gap states
based on STM measurements.\cite{(PbSn)1-xInxTeFualgap_GuanD_PRB15}

Compared with the high pressure technique, the doping method can
not only induce superconductivity in TIs or TCIs but also allow
further experiments to confirm topological superconductivity with
odd parity, such as PCS experiments to observe the ZBP,\cite{CuxBi2Se3ZBCP_Sasaki_PRL11,Sn1-xInxTeTSC_Sasaki_PRL12}
specific heat,\cite{CuxBi2Se3Thermo_YonezawaS_NP17} STM
\cite{CuxBi2Se3_TaoR_PRX18} or NMR \cite{CuxBi2Se3NMR_MatanoK_NP16}
measurements to verify spontaneous RSB/nematicity, and ARPES measurements
to detect gapless surface states.\cite{CuxBi2Se3SS_LahoudE_PRB13,Sn1-xInxTeARPED_Sato_PRL13}
The doped TIs and TCIs hosting superconductivity are summarized in Table.\ref{dopingSC}. 
A sketched phase diagram vs. doping concentration is plotted in Fig.\ref{TItoSC}(a), and
topological phase transitions from a TI/TCI to a TSC are well tunable by doping.
Therefore, Cu$_{x}$Bi$_{2}$Se$_{3}$ and Sn$_{1-x}$In$_{x}$Te
systems are promising TSCs.\cite{TSC_Sato_RPP17}

\section{Artificial Structures}

\subsection{Gate-induced TSCs}

Field-effect transistors (FETs) are the core of the gating technique,
which can help induce superconductivity by modulating the carrier density of 2D materials.\cite{Solidgate_AhnCH_RMP06,2DSC_Saito_NRM17,GateSummary_SaitoY_SST16}
Three types of gating methods are well developed, including those based on
metal-insulator-semiconductor (MIS) FETs,\cite{Solidgate_AhnCH_RMP06}
electric double layer (EDL) FETs \cite{GateSummary_SaitoY_SST16,2DSC_Saito_NRM17}
and solid ion conductor (SIC) FETs.\cite{FeSeSolidion_LeiB_PRB17,FeSeTeSolidion_ZhuCS_PRB17,FeSeSolidion_YingTP_PRL18,LiFeOHFeSeSolidion_MaLK_ArXiv18} The limits of the carrier density in a 2D system modulated by the first
two types are 2$\times$10$^{13}$ cm$^{-2}$ \cite{Solidgate_AhnCH_RMP06} and 1$\times$10$^{15}$ cm$^{-2}$.\cite{Ionliquidgating_Ueno_jpsj14} The SIC FET can contain Li$^{+}$
in the 2D material, and the amount of intercalated Li$^{+}$ can be comparable to
the quantity of the base compound elements.\cite{FeSeSolidion_LeiB_PRB17}
Therefore, these gating techniques can be used to study complex phase diagrams
of 2D materials by controlling their carrier density.

With a gapless, massless and chiral Dirac spectrum,\cite{Graphene_sarma_RMP11,Graphene_Castro_RMP09}
graphene can be tuned to a superconducting state when a `magic' angle exists in
twisted bilayer graphene.\cite{grapheneSC_CaoY_nature18} $T_{c}$ = 1.7 K is observed
when the `magic' angle is 1.1$^{\circ}$. Through the tunability of the carrier density,
the superconducting diagram exhibits similar characteristics to that of cuprates.
Flat bands emerge in the band structure of twisted bilayer graphene, and a Mott-like
insulator phase appears when the Fermi level is tuned at the half-filling of
these flat bands, while two superconducting domes appear adjacent to this half-filling state.\cite{grapheneSC_CaoY_nature18,grapheneSCMott_CaoY_Nature18} These experiments
offer new insights into unconventional superconductors and quantum spin liquids.
Moreover, theoretic predictions have been performed and indicate possible
topological superconductivity in twisted multilayer graphene,\cite{TSCgraphene_XuC_PRL18}
which can host spin triplet $d+id$ order and topologically protected gapless edge states.
Therefore, topological superconductivity in this system is worthy of further investigation.

Much progress has also been made in the FET experiments of topological semimetals.
When the 3D type-II WSM WTe$_{2}$ is fabricated into a monolayer crystal,
the hallmark transport conductance with approximately $e^{2}/h$ per edge is observed
up to 100 K, indicating a quantum spin Hall state in monolayer WTe$_{2}$.\cite{WTe2QSHEmonolayer_WuS_Science18}
If the carrier density in monolayer WTe$_{2}$ is tuned by moderating electrostatic gating,
then a quantum phase transition emerges with increasing gate voltage,
and the quantum spin Hall state can transform into a superconducting state.\cite{WTe2SCmonolayer_Fatemi_Science18,WTe2SCmonolayer_Sajadi_Science18}
The maximum $T_{c}$ is $\sim$ 1 K when the 2D carrier density is modulated to be
more than 1 $\times$10$^{13}$ cm$^{-2}$. The in-plane upper critical field is
much higher than the Pauli limit value. Such high H$_{c2}$ could originate from
various phenomena, such as Ising-type superconductivity,\cite{2DSC_Saito_NRM17}
a reduced electron $g$-factor, a spin-triplet pairing state, or strong spin-orbit scattering.\cite{LayerSC_Klem_PRB75,WTe2SCmonolayer_Fatemi_Science18} The suppression
of the helical edge states by the magnetic field through breaking of the
time reversal symmetry seems to persist in the superconducting state,
which implies nontrivial properties in monolayer WTe$_{2}$,\cite{WTe2SCmonolayer_Sajadi_Science18}
although the pairing state requires more evidence.

Among these three FET methods, the MIS FET technique is the cleanest one,
and only an external electronic field is applied to tune the carrier density,
in contrast with EDL FET and SIC FET techniques. Therefore, in the next subsection
on heterostructures or Josephson junctions, the MIS FET technique is widely used
to control the Fermi level of heterostructures and thus modulate the superconducting state.\cite{review_Beenakker_ARCMP13,MZMsSCsemi_Lutchyn_NRM18} Therefore, gating
techniques are very attractive for exploring TSCs among topological materials.

\subsection{Nanowire/superconductor heterostructures }

\begin{figure}[!thb]
\begin{center}
\includegraphics[width=3.4in]{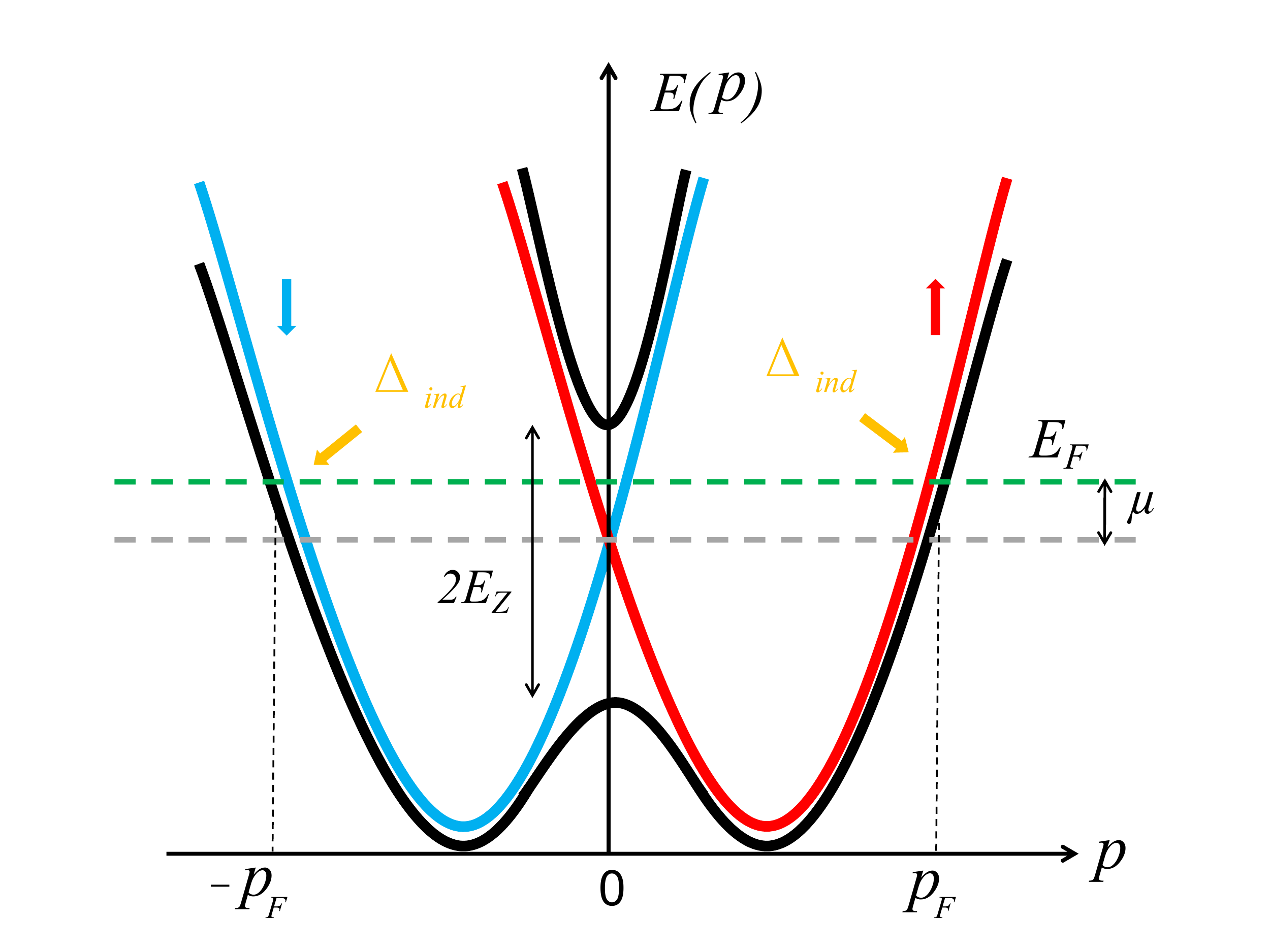}
\end{center}
\caption{\label{Semic-SC} The band evolution of a 1D nanowire coupled to an s-wave superconductor.
First, the degenerate parabolic band of the 1D nanowire can split due to
Rashba SOC. Then, the external magnetic field along the nanowire
causes a Zeeman gap at the crossing point $p = 0$. Moreover, a
superconducting gap $\Delta_{ind}$ is introduced in the nanowire at $p = p_{F}$
by the proximity effect of the s-wave superconductor, which is not shown in the figure.
A different band structure can be obtained by tuning the value of the Zeeman
energy $E_{Z}$.\cite{Al-InAs_DasA_NP12}}
\end{figure}

One effective approach to obtaining TSCs with Majorana
fermions in artificial devices is placing semiconductor nanowires
on an s-wave superconductor. Many similar proposals to detect MZMs in this type of heterostructure have been made. \cite{semicSCnewp_SauJD_PRL10,semicSC_Alicea_PRB10,SemicSc_Lutchyn_PRL10,SemicSc_Oreg_PRL10,review_Beenakker_ARCMP13}
The general experimental setup is as follows. First, the 1D
semiconductor nanowire should exhibit large Rashba SOC, which can
be regarded as a spin-orbital effective field
$\textbf{B}_{so}\propto\textbf{p}\times\textbf{E}$ (where
\textbf{E} is the electric field vertical to the nanowire, while
\textbf{p} is the momentum along the nanowire). This $\textbf{B}_{so}$ is perpendicular
to the nanowire and splits the parabolic band, which results in
crossing of two spin-orbital bands at $p = 0$ (the red band and blue band), as shown in
Fig.\ref{Semic-SC}. Then, if an external magnetic field $B$ is
applied along the axis of the nanowire, a Zeeman gap $2E_{Z}=g\mu_{B}B$
opens at this crossing point, where $g$ is the Land\'{e} $g$-factor and
$\mu_{B}$ is the Bohr magneton. The proximity effect of the s-wave
superconductor and the nanowire will change the band structure by
inducing a superconducting gap $\Delta_{ind}$ at $p = p_{F}$ in the nanowire,
marked by the yellow arrows. The s-wave superconductor can induce
fully gapped superconductivity in the nanowire through the proximity effect.
Then, the Zeeman gap at $p = 0$ can be modulated to control the topological
phase of the nanowire by changing the external magnetic field $B$.
A topological phase transition emerges at $E_{Z}=\sqrt{\Delta_{ind}^{2}+\mu^{2}}$,
where $\mu$ is the chemical potential, as seen in Fig.\ref{Semic-SC}. When $E_{Z}$
is less than $\sqrt{\Delta_{ind}^{2}+\mu^{2}}$, the nanowire is
actually a trivial superconductor similar to the proximate s-wave
superconductor. Subsequently, with increasing $E_{Z}$, the gap at
$p = 0$ will be reduced to zero, and zero-energy bound states
formed by Majorana states will appear at each end of the nanowire. Therefore, if
$E_{Z}>\sqrt{\Delta_{ind}^{2}+\mu^{2}}$, the nanowire enters into
a topological superconducting state.\cite{Al-InAs_DasA_NP12,semicSCnewp_SauJD_PRL10}

Following the above idea, Majorana bound states have been successfully observed in the
heterostructure consisting of normal state (gold)-InSb-superconductor (NbTiN).\cite{InSbSC_Mourik_science12}
Semiconductor InSb nanowires possess strong SOC and a large $g$-factor ($g\approx50$).
These characteristics greatly contribute to modulating the Zeeman gap $2E_{Z}$, which must be
larger than the induced superconducting gap. The induced
superconducting gap here is $\Delta_{ind} \approx$ 0.25 meV, and the
Zeeman gap $E_{Z}/B \approx$ 1.5 meV/T. Therefore, when $B >$ 0.15
T, a topological phase transition will occur because $E_{Z}$ begins
to exceed $\Delta_{ind}$. Majorana bound states have been experimentally
observed at the boundary of the nanowire at $B$ = 0.25 T. By varying the
gate voltage, the chemical potential $\mu$ can be controlled to lower
the critical $B$ at which the topological phase transition emerges.
In addition, with increasing angle between the nanowire and external magnetic field $B$,
the ZBP originating from the Majorana fermion will finally disappear in
the perpendicular case. The amplitude of $dI/dV$ at the ZBP
usually deviates from the theoretical value of 2$e^{2}/h$ at zero
temperature \cite{MF2e2/h_LawKT_PRL09,MF2e2/h_LawKT_PRB10} due to
the finite measurement temperature, where $h$ is the Planck constant
and $e$ is the electric charge.

The observation of MZMs has also been reported in a similar device with an
Al-InAs heterostructure.\cite{Al-InAs_DasA_NP12} InAs is a semiconductor with large SOC
and $g \approx$ 20. A small external magnetic field in this device is required
to observe the ZBPs because the induced superconducting gap $\Delta_{ind} \approx$ 0.05 meV
is very small, which is related to the magnitude of the superconducting gap of the
s-wave superconductor aluminum. In addition, the signature of MZMs can
be detected in several other devices, such as InSb-NbTiN,\cite{InSb-NbTiN_Chur_PRB13}
InSb-Nb \cite{Nb-InSb_Deng_NL12} and InAs-Al heterostructures.\cite{InAs-Al_Albre_Nature16}

To avoid other effects, such as the Kondo effect \cite{KandoEZBP_Gordon_Nat98}
or the so-called `0.7 anomaly',\cite{0.7anomalyZBP_Cron_PRL02,0.7anomaly_Rokh_PRL06}
which could also cause ZBPs in the conductivity measurements of ferromagnetic
atomic chains on an s-wave superconductor, a radiofrequency/microwave technique was
used for a.c. Josephson junctions (a Nb/InSb/Nb device) to detect edge states.\cite{InSb-Nb_Rokhinson_NP12}
The conventional supercurrent is carried by charge 2$e$ Cooper
pairs, with a height of the quantized voltage steps (Shapiro steps) of
$\Delta V=hf_{0}/2e$ in the $I$-$V$ curves, where $f_{0}$ is the
frequency of microwave excitation. When a high magnetic
field is applied, the height of the quantized voltage steps changes to
$hf_{0}/e$, indicating the appearance of charge $e$ quasiparticles
formed by two Majorana fermions in the supercurrent. Therefore,
this doubling of the Shapiro steps is a characteristic signature
of MZMs.

\subsection{TI/superconductor heterostructures }

The proximity effect in TI/superconductor heterostructures has also been
employed as an effective method to induce MZMs.\cite{SCpeonTI_FuL_PRL08}
Instead of a semiconductor nanowire/superconductor heterostructure, a device with a
linear junction between two superconductors mediated by a TI was fabricated,
and a 1D wire of Majorana fermions was engineered along the width of this junction.
Various experiments have been performed to obtain evidence of
Majorana fermions.\cite{Bi2Se3-AlnoMF_Sac_NC11,Bi2Te3-PbnoMF_QuF_SR12,Be2Se3-Pb_YangF_PRB12,Bi2Se3-PbNoMF_Williams_PRL12,Bi2Te3-NbNoMF_Veldh_NM12,Bi2Se3-AlNoMF_ChoS_NC13,HgTe-NbNoMF_Oost_PRX13,Bi2Se3-NbNoMF_Finck_PRX14}
However, detecting the feature of Majorana fermions is quite difficult.
A very important experiment for this type of device is the observation of the
Fraunhofer-type dependence of the critical current $I_{c}$ modulated by the
magnetic field:\cite{Bi2Te3-PbnoMF_QuF_SR12,Be2Se3-Pb_YangF_PRB12}
\begin{equation}\label{LK}
I_{c}(B)=I_{c}(0)|sin\left(\frac{\pi\Phi_{J}}{\Phi_{0}}\right)/\left(\frac{\pi\Phi_{J}}{\Phi_{0}}\right)|,
\end{equation}
where $\Phi_{J}$ is the flux through the effective junction area and $I_{c}(0)$
is the superconducting critical current at zero magnetic field. The usual 2$\pi$-periodic
Josephson supercurrent oscillates with a periodicity of $\Delta B=\Phi_{0}/A$
($A$ is the area of the junction). However, a 4$\pi$-periodic Josephson supercurrent emerges
($\Delta B=2\Phi_{0}/A$), suggesting the existence of MZMs.\cite{HgTe-NbNoMF_Oost_PRX13}

This 4$\pi$-periodic Josephson effect was successfully
obtained in the device fabricated with strained and undoped HgTe on a
CdTe substrate, in which superconductor Nb electrodes were used.\cite{HgTe-NbMF_Wiedenm_NC16}
The Shapiro steps $V_{n}=nhf/2e$ are observed in the $I-V$ curves under high
radiofrequency. The length of the current steps in the Shapiro
steps presents a 4$\pi$-periodic behavior, which may be related to
a p-wave superconductor due to a doublet of topologically
protected gapless Andreev bound states. In a similar device
of HgTe/HgCdTe quantum wells, helical edge states appear in the
supercurrent distributions upon controlling the width of the device,
and this quantum spin Hall effect provides another
insight into TSCs.\cite{HgTe/HgCdTe_Hart_NP14} Subsequently,
another type of HgTe-based Josephson junction was fabricated to
measure the radiofrequency emission spectra. The half Josephson
frequency originating from the 4$\pi$-periodic gapless Andreev
doublet around zero energy was confirmed in this device, although
the coherence time of 0.3-4 ns was very short.\cite{HgTe-AlMF_Deacon_PRX17}

In another experiment, a superconductor was coupled to a
thin film of a quantum anomalous Hall insulator to detect the chiral
Majorana fermion modes, i.e., a superconductor Nb bar was fabricated
on a (Cr$_{0.12}$Bi$_{0.26}$Sb$_{0.62}$)$_{2}$Te$_{3}$ film.\cite{QAHE-SC_HeQL_Science17}
An external magnetic field was applied to modulate the topological
state of the device, and a half-integer plateau of the longitudinal conductance
($e^{2}/2h$) was obtained. This unique transport signature
indicates the existence of MZMs. Therefore, this type of device may become
a promising component in topological quantum computation.

ARPES and STM are both powerful techniques to detect the MZMs in
TI films grown on superconductors.\cite{Be2Se3MF_WangE_NP13,review_Beenakker_ARCMP13,Bi2Te3-NbSe2MF_XuJP_PRL15}
A straightforward approach is to grow Bi$_{2}$Se$_{3}$ thin films by molecular
beam epitaxy (MBE) on the s-wave superconductor NbSe$_{2}$. Experimentally, a
superconducting gap is observed in the Bi$_{2}$Se$_{3}$ film, coexisting with
topological surface states, which may generate topological superconductivity in the TI
film.\cite{Bi2Se3-NbSe2_WangMX_sience12} When Bi$_{2}$Se$_{3}$ films are grown
on the d-wave superconductor Bi$_{2}$Sr$_{2}$CaCu$_{2}$O$_{8+\delta}$, fully gapped
topological surface states are detected by ARPES.\cite{Be2Se3MF_WangE_NP13}
This isotropic superconducting gap structure is different from the bulk superconducting gap
structure and may provide another route to realizing MZMs. Moreover, MZMs located
in the vortex core in the Bi$_{2}$Te$_{3}$/NbSe$_{2}$ heterostructure have been
successfully observed by STM measurements.\cite{Bi2Te3-NbSe2MF_XuJP_PRL15,Bi2Te3-NbSe2_SunHH_PRL16}
To conclude, TI/superconductor heterostructures also offer a good device platform
to realize realistic topological quantum computation.

\subsection{Natural heterostructures}

In addition to artificial heterostructures, some attempts have been
made to synthesize natural heterostructures to realize TSCs. For such natural
heterostructures, the samples are usually bulk single crystals.
A good example of natural heterostructures whose TI features have been
experimentally confirmed is (PbSe)$_{5}$(Bi$_{2}$Se$_{3}$)$_{3m}$ ($m$ =1, 2, ...).\cite{(PbSe)5(Bi2Se3)6_Naka_PRL12}
As mentioned in the previous section, this compound can be viewed as the alternation of
(PbSe)$_{5}$ layer units and (Bi$_{2}$Se$_{3}$)$_{3}$ layer units.
Bi$_{2}$Se$_{3}$ is a TI, and PbSe is an ordinary insulator. The inversion
symmetry breaking at the interface of these two units will induce Rashba
splitting of the topological surface state at $m$ = 2. With increasing $m$, a
topological phase transition occurs from a trivial phase at $m$=1
to a TI phase at $m \geq$2. When Cu is intercalated in the Bi$_{2}$Se$_{3}$
units of this compound, forming Cu$_{x}$(PbSe)$_{5}$(Bi$_{2}$Se$_{3}$)$_{6}$,
a superconducting transition emerges when the value $x$ is between 0.3 and 2.5.\cite{Cux(PbSe)5(Bi2Se3)6_Sasaki_PRB14}
Unconventional superconductivity with a nodal gap and nematic superconductivity
have also been observed, as mentioned in subsection A of section IV. Therefore,
Cu$_{x}$(PbSe)$_{5}$(Bi$_{2}$Se$_{3}$)$_{6}$ may be a candidate
for TSCs with a nodal gap.\cite{TSC_Sato_RPP17} Although superconductivity
exists in the Ag-doped (PbSe)$_{5}$(Bi$_{2}$Se$_{3}$)$_{3}$ ($m$=1) samples, with
$T_{c}$ = 1.7 K,\cite{Agdoped(PbSe)5(Bi2Se3)3SC_FangL_PRB14} no superconducting
transition appears in the Ag-doped TI samples (PbSe)$_{5}$(Bi$_{2}$Se$_{3}$)$_{6}$ ($m$=2).

Other natural structures, such as misfit layer compounds, also exist.\cite{Misfit_Wiegers_PSSC96}
Misfit compounds are a large family characterized by (MX)$_{1+y}$(TX$_{2}$)$_{n}$ ($y$ = 0.08-0.28,
$n$ = 1, 2, 3), where M = Sn, Pb, Sb, Bi or a lanthanide; X = S,
Se or Te; and T = Ti, V, Cr, Nb or Ta. Among these compounds, the
TX$_{2}$ layer is usually a superconducting unit, and the MX layer
can be regarded as a modulated structural unit. This interesting
natural heterostructure could possess an abnormally large upper
critical field,\cite{MisfitHc2_BaiH_JPCM18} which may
originate from multiband effects or similar spin-valley locking
in the 2D limit case.\cite{2DSC_Saito_NRM17} The misfit compounds
exhibit inversion symmetry breaking at the interface of the
heterostructure, which may be accompanied by the Rashba effect in
a strong SOC system, such as (PbSe)$_{5}$(Bi$_{2}$Se$_{3}$)$_{6}$. These
natural heterostructures lower the dimensionality of crystals,
which could contribute to a large anisotropic upper critical field,
such as in the 2D limit case of ion-gated MoS$_{2}$
\cite{MoS2SVK_Saito_NP16} and NbSe$_{2}$ \cite{NbSe22D_XiX_NP16}
bilayers. If the MX layer is composed of topological
materials, such as the predicted monolayer TCIs SnTe, PbS, PbSe, and PbTe,\cite{SnTeTheory_Hsieh_NC12,PbSe2DTCI_Wrasse_NL14,IV-VITCI_LiuJ_NanoL15,SnTe2D_LiuJ_NM14,PbS2DTCI_WanWh_AM17}
the proximity effect of the superconducting TX$_{2}$ layer may
induce Majorana bound states \cite{SCpeonTI_FuL_PRL08} in the
adjacent surface states of the TI layers, similar to those in the artificial
TI/superconductor heterostructures mentioned in the previous
subsection. Unfortunately, no experiments confirming the
topological features in these misfit compounds have been reported.

\subsection{Other artificial structures}

Moreover, several other heterostructures to realize MZMs have been studied.
The heterostructure of ferromagnetic atomic chains on an s-wave superconductor
has been successfully fabricated, and ZBPs can be observed in these devices,
which may be caused by the Kondo effect or disorder.\cite{ZBPKondo_Lee_PRL12,InSb-NbTiN_Chur_PRB13,ZPBdisorder_LiuJ_PRL12,ZBPkondo_Pikulin_NPJ12,ZBPkondo_Kells_PRB12}  ZBPs at the boundary of the ferromagnetic atomic chain can be successfully
detected in the artificial structure of magnetic atomic chains on the surface
of an s-wave superconductor by means of a high-resolution spectroscopic imaging technique.\cite{Fe-PbZBP_Nadj_Science14}

In other artificial structures, when quantum Hall edges are coupled with
a superconductor, new excitations, such as non-Abelian anyons, will emerge,
and nonlocal transport could possibly be used to detect Majorana fermions.\cite{Zeromode_Clarke_NatP14}
Subsequently, graphene/superconductor heterostructures have drawn attention
because of the existence of a quantum Hall state in graphene. SOC in this junction
is not necessary, while strong SOC is required in semiconductor or TI/superconductor
junctions.\cite{MZMinGraphene_SanJP_PRX15} Topologically protected Majorana bound state
exists and may be detected by Andreev spectroscopy and Fraunhofer pattern anomalies.
Although many endeavors have been made,\cite{MZMinGraphene_SanJP_PRX15} further studies
to realize topological superconductivity are expected.

\section{Summary and Outlook}

To summarize, research to discover and verify TSCs has become one of the most active
fields in condensed matter physics due to the potential applications in
fault-tolerant topological quantum computation.\cite{topologicalquantumcomput_Nayak_RMP08}
Decades ago, the TSC candidates Sr$_{2}$RuO$_{4}$ \cite{TSC_Sato_RPP17,ChiralSC_Kallin_RPP16,Sr2RuO4_MaenoY_JPCS12}
and UPt$_{3}$ \cite{TSC_Sato_RPP17,ChiralSC_Kallin_RPP16,UPt3_Joynt_RMP02}
were inadvertently discovered. However, either the chiral edge states or the MZMs in
Sr$_{2}$RuO$_{4}$ have not been detected, and little evidence
to support the spin-triplet pairing state for UPt$_{3}$ has been found.
Subsequently, research on topological materials became a hot topic,
which opened up new horizons for exploring TSCs.\cite{TIandTSC_QiXL_RMP11,TI_Hasan_RMP10}
High pressure application is a relatively powerful method to induce superconductivity in
TIs and semimetals, but the disadvantage of
this technique is the difficulty of conducting further experiments to
detect the pairing symmetry and gapless surface state.
Therefore, the hard-tip contact method seems to provide a new approach
to inducing superconductivity and simultaneously detecting the ZBP
of gapless surface states.

The doping method offers an effective platform to study
superconductivity in topological materials, and most experiments
reveal the unique features associated with TSCs, such as
spontaneous RSB, ZBPs,\cite{CuxBi2Se3NMR_MatanoK_NP16,CuxBi2Se3Thermo_YonezawaS_NP17}
and MZMs.\cite{CuxBi2Se3ZBCP_Sasaki_PRL11} One of the application
goals in studying TSCs is to realize topological quantum computation,
and thus, many efforts have been made to fabricate devices.
In this direction, the most progress has been made in probing MZMs
in artificial heterostructures.\cite{MZMsSCsemi_Lutchyn_NRM18,review_Beenakker_ARCMP13,TSC_Sato_RPP17}
In addition, more approaches to finding TSCs, such as
natural heterostructures,\cite{Cux(PbSe)5(Bi2Se3)6_Sasaki_PRB14}
gating-induced TSCs,\cite{WTe2SCmonolayer_Fatemi_Science18,WTe2SCmonolayer_Sajadi_Science18}
or iron-based superconductors with nontrivial band
structures,\cite{FeSeTeTSC_ZhangP_Science18} are encouraged. The material systems
with coexistence of topological bands and superconductivity, such as
the layered compound PbTaSe$_{2}$,\cite{PbTaSe2_BianG_NatC,PbTaSe2_GuanSY_SciAdv16},
have become promising candidate systems for TSCs.
Recently, thousands of topological (crystalline) insulators and
topological semimetals have been theoretically classified.\cite{CatalogueTM_ZhangTT_nature19,TM_TangF_nature19,TM_VergnioryMG_nature19}
We believe that rapid development in the field of topological
materials will greatly promote the investigation of topological
superconductivity and motivate novel devices.

\section{ACKNOWLEDGMENTS}

\noindent We thank Yi Zhou, Xin Lu, Yi Zheng, Guanghan Cao, and
Fuchun Zhang for helpful discussions. This work was supported by
the National Key R\&D Program of China (Grant No. 2016YFA0300402)
and the National Science Foundation of China (Grant No. 11774305).


\begin{thebibliography}{100}

\bibitem{topologicalquantumcomput_Nayak_RMP08}
C.~{Nayak}, S.~H. {Simon}, A.~{Stern}, M.~{Freedman}, S.~{Das Sarma},
  \emph{Rev. Mod. Phys.} \textbf{2008}, \emph{80}, 1083.

\bibitem{HeliumDroplet_Volovik_Oxford03}
G.~E. {Volovik}, \emph{(Oxford: Oxford University Press)} \textbf{2003}.

\bibitem{Paired_ReadN_PRB00}
N.~Read, D.~Green, \emph{Phys. Rev. B} \textbf{2000}, \emph{61}, 10267.

\bibitem{quantumwire_Kitaev_pu01}
A.~Y. {Kitaev}, \emph{Phys. Usp.} \textbf{2001}, \emph{44}, 131.

\bibitem{MajoranaF_Majorana_Il1937}
E.~{Majorana}, \emph{Il Nuovo Cimento} \textbf{1937}, \emph{14}, 171.

\bibitem{Majorana_Wilczek_NatPhys}
F.~Wilczek, \emph{Nat. Phys.} \textbf{2009}, \emph{5}, 614.

\bibitem{review_Beenakker_ARCMP13}
C.~W.~J. {Beenakker}, \emph{Annu. Rev. Condens. Matter Phys.} \textbf{2013},
  \emph{4}, 113.

\bibitem{TI3D_FuL_PRL07}
L.~Fu, C.~L. Kane, E.~J. Mele, \emph{Phys. Rev. Lett.} \textbf{2007},
  \emph{98}, 106803.

\bibitem{TIs_Ful_PRB07}
L.~Fu, C.~L. Kane, \emph{Phys. Rev. B} \textbf{2007}, \emph{76}, 045302.

\bibitem{SCpeonTI_FuL_PRL08}
L.~Fu, C.~L. Kane, \emph{Phys. Rev. Lett.} \textbf{2008}, \emph{100}, 096407.

\bibitem{TIandTSC_QiXL_RMP11}
X.-L. Qi, S.-C. Zhang, \emph{Rev. Mod. Phys.} \textbf{2011}, \emph{83}, 1057.

\bibitem{CuxBi2Se3Theory_FuL_PRL10}
L.~Fu, E.~Berg, \emph{Phys. Rev. Lett.} \textbf{2010}, \emph{105}, 097001.

\bibitem{semicSCnewp_SauJD_PRL10}
J.~D. Sau, R.~M. Lutchyn, S.~Tewari, S.~Das~Sarma, \emph{Phys. Rev. Lett.}
  \textbf{2010}, \emph{104}, 040502.

\bibitem{TSC_Sato_RPP17}
M.~{Sato}, Y.~{Ando}, \emph{Rep. Prog. Phys.} \textbf{2017}, \emph{80}, 076501.

\bibitem{TCIandTSC_Ando_ARCMP15}
Y.~{Ando}, L.~{Fu}, \emph{Annu. Rev. Condens. Matter Phys.} \textbf{2015},
  \emph{6}, 361.

\bibitem{Sr2RuO4NMR_Ishida_Nature98}
K.~{Ishida}, H.~{Mukuda}, Y.~{Kitaoka}, K.~{Asayama}, Z.~Q. {Mao}, Y.~{Mori},
  Y.~{Maeno}, \emph{Nature} \textbf{1998}, \emph{396}, 658.

\bibitem{Sr2RuO4USR_Luke_Nature98}
G.~M. {Luke}, Y.~{Fudamoto}, K.~M. {Kojima}, M.~I. {Larkin}, J.~{Merrin},
  B.~{Nachumi}, Y.~J. {Uemura}, Y.~{Maeno}, Z.~Q. {Mao}, Y.~{Mori},
  H.~{Nakamura}, M.~{Sigrist}, \emph{Nature} \textbf{1998}, \emph{394}, 558.

\bibitem{UPt3uSR_Luke_PRL93}
G.~M. Luke, A.~Keren, L.~P. Le, W.~D. Wu, Y.~J. Uemura, D.~A. Bonn,
  L.~Taillefer, J.~D. Garrett, \emph{Phys. Rev. Lett.} \textbf{1993},
  \emph{71}, 1466.

\bibitem{InSbSC_Mourik_science12}
V.~{Mourik}, K.~{Zuo}, S.~M. {Frolov}, S.~R. {Plissard}, E.~P.~A.~M. {Bakkers},
  L.~P. {Kouwenhoven}, \emph{Science} \textbf{2012}, \emph{336}, 1003.

\bibitem{Bi2Te3-NbSe2MF_XuJP_PRL15}
J.-P. {Xu}, M.-X. {Wang}, Z.~L. {Liu}, J.-F. {Ge}, X.~{Yang}, C.~{Liu}, Z.~A.
  {Xu}, D.~{Guan}, C.~L. {Gao}, D.~{Qian}, Y.~{Liu}, Q.-H. {Wang}, F.-C.
  {Zhang}, Q.-K. {Xue}, J.-F. {Jia}, \emph{Phys. Rev. Lett.} \textbf{2015},
  \emph{114}, 017001.

\bibitem{Bi2Te3-NbSe2_SunHH_PRL16}
H.-H. {Sun}, K.-W. {Zhang}, L.-H. {Hu}, C.~{Li}, G.-Y. {Wang}, H.-Y. {Ma},
  Z.-A. {Xu}, C.-L. {Gao}, D.-D. {Guan}, Y.-Y. {Li}, C.~{Liu}, D.~{Qian},
  Y.~{Zhou}, L.~{Fu}, S.-C. {Li}, F.-C. {Zhang}, J.-F. {Jia}, \emph{Phys. Rev.
  Lett.} \textbf{2016}, \emph{116}, 257003.

\bibitem{HgTe-NbMF_Wiedenm_NC16}
J.~{Wiedenmann}, E.~{Bocquillon}, R.~S. {Deacon}, S.~{Hartinger},
  O.~{Herrmann}, T.~M. {Klapwijk}, L.~{Maier}, C.~{Ames}, C.~{Br{\"u}ne},
  C.~{Gould}, A.~{Oiwa}, K.~{Ishibashi}, S.~{Tarucha}, H.~{Buhmann}, L.~W.
  {Molenkamp}, \emph{Nat. Commun.} \textbf{2016}, \emph{7}, 10303.

\bibitem{HgTe-AlMF_Deacon_PRX17}
R.~S. Deacon, J.~Wiedenmann, E.~Bocquillon, F.~Dom\'{\i}nguez, T.~M. Klapwijk,
  P.~Leubner, C.~Br\"une, E.~M. Hankiewicz, S.~Tarucha, K.~Ishibashi,
  H.~Buhmann, L.~W. Molenkamp, \emph{Phys. Rev. X} \textbf{2017}, \emph{7},
  021011.

\bibitem{HgTe/HgCdTe_Hart_NP14}
S.~{Hart}, H.~{Ren}, T.~{Wagner}, P.~{Leubner}, M.~{M{\"u}hlbauer},
  C.~{Br{\"u}ne}, H.~{Buhmann}, L.~W. {Molenkamp}, A.~{Yacoby}, \emph{Nat.
  Phys.} \textbf{2014}, \emph{10}, 638.

\bibitem{superconductivity_JFAnnett_book04}
J.~F. Annett, \emph{(Oxford Master Series in Condensed Matter Physics, Oxford,
  2004)} .

\bibitem{CuprateSC_Tsuei_RMP00}
C.~C. Tsuei, J.~R. Kirtley, \emph{Rev. Mod. Phys.} \textbf{2000}, \emph{72},
  969.

\bibitem{IronbasedSC_Stewart_RMP11}
G.~R. Stewart, \emph{Rev. Mod. Phys.} \textbf{2011}, \emph{83}, 1589.

\bibitem{Sr2RuO4_MaenoY_JPCS12}
Y.~{Maeno}, S.~{Kittaka}, T.~{Nomura}, S.~{Yonezawa}, K.~{Ishida}, \emph{J.
  Phys. Soc. Jpn.} \textbf{2012}, \emph{81}, 011009.

\bibitem{CrBasedSC_CaoGH_PM17}
G.-H. Cao, J.-K. Bao, Z.-T. Tang, Y.~Liu, H.~Jiang, \emph{Philos. Mag.}
  \textbf{2017}, \emph{97}, 591.

\bibitem{La2BaxCuO4_Bednorz_ZPB86}
J.~G. {Bednorz}, K.~A. {M{\"u}ller}, \emph{Z. Phys. B} \textbf{1986},
  \emph{64}, 189.

\bibitem{Sr2RuO4_Maeno_Nature94}
Y.~{Maeno}, H.~{Hashimoto}, K.~{Yoshida}, S.~{Nishizaki}, T.~{Fujita}, J.~G.
  {Bednorz}, F.~{Lichtenberg}, \emph{Nature} \textbf{1994}, \emph{372}, 532.

\bibitem{Sr2RuO4_Mack_RMP03}
A.~P. Mackenzie, Y.~Maeno, \emph{Rev. Mod. Phys.} \textbf{2003}, \emph{75},
  657.

\bibitem{PauliLimit_Clogston_PRL62}
A.~M. Clogston, \emph{Phys. Rev. Lett.} \textbf{1962}, \emph{9}, 266.

\bibitem{Sr2RuO4PS_Nelson_Science04}
K.~D. {Nelson}, Z.~Q. {Mao}, Y.~{Maeno}, Y.~{Liu}, \emph{Science}
  \textbf{2004}, \emph{306}, 1151.

\bibitem{IcJunction_Geshk_PRB87}
V.~B. Geshkenbein, A.~I. Larkin, A.~Barone, \emph{Phys. Rev. B} \textbf{1987},
  \emph{36}, 235.

\bibitem{ChiralSC_Kallin_RPP16}
C.~{Kallin}, J.~{Berlinsky}, \emph{Rep. Prog. Phys.} \textbf{2016}, \emph{79},
  054502.

\bibitem{Sr2RuO4Ic_Kidw_Science06}
F.~{Kidwingira}, J.~D. {Strand}, D.~J. {Van Harlingen}, Y.~{Maeno},
  \emph{Science} \textbf{2006}, \emph{314}, 1267.

\bibitem{Halfvortex_JangJ_Science11}
J.~{Jang}, D.~G. {Ferguson}, V.~{Vakaryuk}, R.~{Budakian}, S.~B. {Chung}, P.~M.
  {Goldbart}, Y.~{Maeno}, \emph{Science} \textbf{2011}, \emph{331}, 186.

\bibitem{HalfVortexbyMZMs_Kopnin_PRB91}
N.~B. Kopnin, M.~M. Salomaa, \emph{Phys. Rev. B} \textbf{1991}, \emph{44},
  9667.

\bibitem{Sr2RuO4edgestate_Kash_PRL11}
S.~{Kashiwaya}, H.~{Kashiwaya}, H.~{Kambara}, T.~{Furuta}, H.~{Yaguchi},
  Y.~{Tanaka}, Y.~{Maeno}, \emph{Phys. Rev. Lett.} \textbf{2011}, \emph{107},
  077003.

\bibitem{Sr2RuO4_Kallin_JPCM09}
C.~{Kallin}, A.~J. {Berlinsky}, \emph{J. Phys.-Condes. Matter} \textbf{2009},
  \emph{21}, 164210.

\bibitem{Sr2RuO4_Pustogow_arXiv19}
A.~{Pustogow}, Y.~{Luo}, A.~{Chronister}, Y.~S. {Su}, D.~A. {Sokolov},
  F.~{Jerzembeck}, A.~P. {Mackenzie}, C.~W. {Hicks}, N.~{Kikugawa}, S.~{Raghu},
  E.~D. {Bauer}, S.~E. {Brown}, \emph{arXiv:1904.00047} .

\bibitem{TopoSurfSinNodeSC_Schnyd_JPCM15}
A.~P. {Schnyder}, P.~M.~R. {Brydon}, \emph{J. Phys.-Condes. Matter}
  \textbf{2015}, \emph{27}, 243201.

\bibitem{Sr2RuO4MF_Yuji_PRL13}
Y.~Ueno, A.~Yamakage, Y.~Tanaka, M.~Sato, \emph{Phys. Rev. Lett.}
  \textbf{2013}, \emph{111}, 087002.

\bibitem{UPt3_Joynt_RMP02}
R.~Joynt, L.~Taillefer, \emph{Rev. Mod. Phys.} \textbf{2002}, \emph{74}, 235.

\bibitem{UPt3Kerreffect_Schemm_science14}
E.~R. {Schemm}, W.~J. {Gannon}, C.~M. {Wishne}, W.~P. {Halperin},
  A.~{Kapitulnik}, \emph{Science} \textbf{2014}, \emph{345}, 190.

\bibitem{UPt3fwave_Tsuts_JSPS13}
Y.~{Tsutsumi}, M.~{Ishikawa}, T.~{Kawakami}, T.~{Mizushima}, M.~{Sato},
  M.~{Ichioka}, K.~{Machida}, \emph{J. Phys. Soc. Jap.} \textbf{2013},
  \emph{82}, 113707.

\bibitem{UPt3Twofold_Machid_PRL12}
Y.~Machida, A.~Itoh, Y.~So, K.~Izawa, Y.~Haga, E.~Yamamoto, N.~Kimura,
  Y.~Onuki, Y.~Tsutsumi, K.~Machida, \emph{Phys. Rev. Lett.} \textbf{2012},
  \emph{108}, 157002.

\bibitem{UPt3GapS_Strand_Science10}
J.~D. {Strand}, D.~J. {Bahr}, D.~J. {Van Harlingen}, J.~P. {Davis}, W.~J.
  {Gannon}, W.~P. {Halperin}, \emph{Science} \textbf{2010}, \emph{328}, 1368.

\bibitem{UPt3TSC_Tsutsumi_JSPS13}
Y.~{Tsutsumi}, M.~{Ishikawa}, T.~{Kawakami}, T.~{Mizushima}, M.~{Sato},
  M.~{Ichioka}, K.~{Machida}, \emph{J. Phys. Soc. Jpn} \textbf{2013},
  \emph{82}, 113707.

\bibitem{UPt3TSCTheory_Shingo_PRB16}
S.~Kobayashi, Y.~Yanase, M.~Sato, \emph{Phys. Rev. B} \textbf{2016}, \emph{94},
  134512.

\bibitem{IronBasedSC}
G.~R. Stewart, \emph{Rev. Mod. Phys.} \textbf{2011}, \emph{83}, 1589.

\bibitem{TopologyinIronSC_HaoN_arXiv18}
N.~{Hao}, J.~{Hu}, \emph{Natl. Sci. Rev.} \textbf{2019}, \emph{6}, 213.

\bibitem{FeSeToP_HaoNN_PRX14}
N.~Hao, J.~Hu, \emph{Phys. Rev. X} \textbf{2014}, \emph{4}, 031053.

\bibitem{FeSeSrTiO3SC_WangQY_CPL12}
Q.-Y. {Wang}, Z.~{Li}, W.-H. {Zhang}, Z.-C. {Zhang}, J.-S. {Zhang}, W.~{Li},
  H.~{Ding}, Y.-B. {Ou}, P.~{Deng}, K.~{Chang}, J.~{Wen}, C.-L. {Song},
  K.~{He}, J.-F. {Jia}, S.-H. {Ji}, Y.-Y. {Wang}, L.-L. {Wang}, X.~{Chen},
  X.-C. {Ma}, Q.-K. {Xue}, \emph{Chin. Phys. Lett.} \textbf{2012}, \emph{29},
  037402.

\bibitem{TopEdgeStFeSeSTO_WangZF_NM16}
Z.~F. {Wang}, H.~{Zhang}, D.~{Liu}, C.~{Liu}, C.~{Tang}, C.~{Song}, Y.~{Zhong},
  J.~{Peng}, F.~{Li}, C.~{Nie}, L.~{Wang}, X.~J. {Zhou}, X.~{Ma}, Q.~K. {Xue},
  F.~{Liu}, \emph{Nat. Mater.} \textbf{2016}, \emph{15}, 968.

\bibitem{FeTeSe-STOTopology_ShiX_SB17}
X.~{Shi}, Z.~{Han}, P.~{Richard}, X.~{Wu}, X.~{Peng}, T.~{Qian}, S.~{Wang},
  J.~{Hu}, Y.~{Sun}, H.~{Ding}, \emph{Sci. Bull.} \textbf{2017}, \emph{62},
  503.

\bibitem{FeTeSeFilm_WuXX_PRB16}
X.~{Wu}, S.~{Qin}, Y.~{Liang}, H.~{Fan}, J.~{Hu}, \emph{Phys. Rev. B}
  \textbf{2016}, \emph{93}, 115129.

\bibitem{FeTeSeMZMs_WangD_Science18}
D.~{Wang}, L.~{Kong}, P.~{Fan}, H.~{Chen}, S.~{Zhu}, W.~{Liu}, L.~{Cao},
  Y.~{Sun}, S.~{Du}, J.~{Schneeloch}, R.~{Zhong}, G.~{Gu}, L.~{Fu}, H.~{Ding},
  H.-J. {Gao}, \emph{Science} \textbf{2018}, \emph{362}, 333.

\bibitem{FeTe0.55Se0.45CdGM_ChenM_NC18}
M.~{Chen}, X.~{Chen}, H.~{Yang}, Z.~{Du}, X.~{Zhu}, E.~{Wang}, H.-H. {Wen},
  \emph{Nat. Commun.} \textbf{2018}, \emph{9}, 970.

\bibitem{LiFeOHFeSeSC_LuXF_NM15}
X.~F. {Lu}, N.~Z. {Wang}, H.~{Wu}, Y.~P. {Wu}, D.~{Zhao}, X.~Z. {Zeng}, X.~G.
  {Luo}, T.~{Wu}, W.~{Bao}, G.~H. {Zhang}, F.~Q. {Huang}, Q.~Z. {Huang}, X.~H.
  {Chen}, \emph{Nat. Mater.} \textbf{2015}, \emph{14}, 325.

\bibitem{FeSe_HsuFC_PNAS08}
F.-C. {Hsu}, J.-Y. {Luo}, K.-W. {Yeh}, T.-K. {Chen}, T.-W. {Huang}, P.~M. {Wu},
  Y.-C. {Lee}, Y.-L. {Huang}, Y.-Y. {Chu}, D.-C. {Yan}, M.-K. {Wu}, \emph{Proc.
  Natl. Acad. Sci. U.S.A.} \textbf{2008}, \emph{105}, 14262.

\bibitem{LiFeOHFeSeFullygap_Smidman_PRB17}
M.~Smidman, G.~M. Pang, H.~X. Zhou, N.~Z. Wang, W.~Xie, Z.~F. Weng, Y.~Chen,
  X.~L. Dong, X.~H. Chen, Z.~X. Zhao, H.~Q. Yuan, \emph{Phys. Rev. B}
  \textbf{2017}, \emph{96}, 014504.

\bibitem{LiFeOHFeSeFullygapSTM_NiuXH_PRB15}
X.~H. Niu, R.~Peng, H.~C. Xu, Y.~J. Yan, J.~Jiang, D.~F. Xu, T.~L. Yu, Q.~Song,
  Z.~C. Huang, Y.~X. Wang, B.~P. Xie, X.~F. Lu, N.~Z. Wang, X.~H. Chen, Z.~Sun,
  D.~L. Feng, \emph{Phys. Rev. B} \textbf{2015}, \emph{92}, 060504.

\bibitem{LiFeOHFeSeMZM_LiuQ_PRX18}
Q.~{Liu}, C.~{Chen}, T.~{Zhang}, R.~{Peng}, Y.-J. {Yan}, C.-H.-P. {Wen},
  X.~{Lou}, Y.-L. {Huang}, J.-P. {Tian}, X.-L. {Dong}, G.-W. {Wang}, W.-C.
  {Bao}, Q.-H. {Wang}, Z.-P. {Yin}, Z.-X. {Zhao}, D.-L. {Feng}, \emph{Phys.
  Rev. X} \textbf{2018}, \emph{8}, 041056.

\bibitem{CdGMS-wave_Caroli_PL64}
C.~{Caroli}, P.~G. {De Gennes}, J.~{Matricon}, \emph{Phys. Lett.}
  \textbf{1964}, \emph{9}, 307.

\bibitem{CdGMP-wave_Volovik_TJP96}
G.~E. {Volovik}, \emph{Turk. J. Phys.} \textbf{1996}, \emph{20}, 697.

\bibitem{LiFeOHFeSeZn_DuZ_NP18}
Z.~{Du}, X.~{Yang}, D.~{Altenfeld}, Q.~{Gu}, H.~{Yang}, I.~{Eremin}, P.~J.
  {Hirschfeld}, I.~I. {Mazin}, H.~{Lin}, X.~{Zhu}, H.-H. {Wen}, \emph{Nat.
  Phys.} \textbf{2018}, \emph{14}, 134.

\bibitem{GapinFeBasedSC_Hirschf_RPP11}
P.~J. {Hirschfeld}, M.~M. {Korshunov}, I.~I. {Mazin}, \emph{Rep. Prog. Phys.}
  \textbf{2011}, \emph{74}, 124508.

\bibitem{LiFeAsTSC_ZhangP_arXiv18}
P.~{Zhang}, X.~{Wu}, K.~{Yaji}, G.~{Dai}, X.~{Wang}, C.~{Jin}, J.~{Hu},
  R.~{Thomale}, T.~{Kondo}, S.~{Shin}, \emph{arXiv:1803.00846} .

\bibitem{CaFeAs2TSC_WuXX_PRB15}
X.~{Wu}, S.~{Qin}, Y.~{Liang}, C.~{Le}, H.~{Fan}, J.~{Hu}, \emph{Phys. Rev. B}
  \textbf{2015}, \emph{91}, 081111.

\bibitem{K2Cr3As3_BaoJK_PRX15}
J.-K. {Bao}, J.-Y. {Liu}, C.-W. {Ma}, Z.-H. {Meng}, Z.-T. {Tang}, Y.-L. {Sun},
  H.-F. {Zhai}, H.~{Jiang}, H.~{Bai}, C.-M. {Feng}, Z.-A. {Xu}, G.-H. {Cao},
  \emph{Phys. Rev. X} \textbf{2015}, \emph{5}, 011013.

\bibitem{Rb2Cr3As3_TangZT_PRB15}
Z.-T. {Tang}, J.-K. {Bao}, Y.~{Liu}, Y.-L. {Sun}, A.~{Ablimit}, H.-F. {Zhai},
  H.~{Jiang}, C.-M. {Feng}, Z.-A. {Xu}, G.-H. {Cao}, \emph{Phys. Rev. B}
  \textbf{2015}, \emph{91}, 020506.

\bibitem{Cs2Cr3As3_TangZT_SCM15}
Z.-T. {Tang}, J.-K. {Bao}, Z.~{Wang}, H.~{Bai}, H.~{Jiang}, Y.~{Liu}, H.-F.
  {Zhai}, C.-M. {Feng}, Z.-A. {Xu}, G.-H. {Cao}, \emph{Sci. China-Mater.}
  \textbf{2015}, \emph{58}, 16.

\bibitem{Na2Cr3As3_MuQG_PRM18}
Q.-G. {Mu}, B.-B. {Ruan}, B.-J. {Pan}, T.~{Liu}, J.~{Yu}, K.~{Zhao}, G.-F.
  {Chen}, Z.-A. {Ren}, \emph{Phys. Rev. Materials} \textbf{2018}, \emph{2},
  034803.

\bibitem{K2Cr3As3Hc2_Balak_PRB15}
F.~F. Balakirev, T.~Kong, M.~Jaime, R.~D. McDonald, C.~H. Mielke, A.~Gurevich,
  P.~C. Canfield, S.~L. Bud'ko, \emph{Phys. Rev. B} \textbf{2015}, \emph{91},
  220505.

\bibitem{K2Cr3As3Hc2_ZuoHK_PRB17}
H.~{Zuo}, J.-K. {Bao}, Y.~{Liu}, J.~{Wang}, Z.~{Jin}, Z.~{Xia}, L.~{Li},
  Z.~{Xu}, J.~{Kang}, Z.~{Zhu}, G.-H. {Cao}, \emph{Phys. Rev. B} \textbf{2017},
  \emph{95}, 014502.

\bibitem{Rb2Cr3As3NMR_YangJ_PRL15}
J.~Yang, Z.~T. Tang, G.~H. Cao, G.-Q. Zheng, \emph{Phys. Rev. Lett.}
  \textbf{2015}, \emph{115}, 147002.

\bibitem{K2Cr3As3NMR_ZhiHZ_PRL15}
H.~Z. Zhi, T.~Imai, F.~L. Ning, J.-K. Bao, G.-H. Cao, \emph{Phys. Rev. Lett.}
  \textbf{2015}, \emph{114}, 147004.

\bibitem{K2Cr3As3USR_Adroja_PRB15}
D.~T. Adroja, A.~Bhattacharyya, M.~Telling, Y.~Feng, M.~Smidman, B.~Pan,
  J.~Zhao, A.~D. Hillier, F.~L. Pratt, A.~M. Strydom, \emph{Phys. Rev. B}
  \textbf{2015}, \emph{92}, 134505.

\bibitem{K2Cr3As3node_PangGM_PRB}
G.~M. Pang, M.~Smidman, W.~B. Jiang, J.~K. Bao, Z.~F. Weng, Y.~F. Wang,
  L.~Jiao, J.~L. Zhang, G.~H. Cao, H.~Q. Yuan, \emph{Phys. Rev. B}
  \textbf{2015}, \emph{91}, 220502.

\bibitem{DAChighpressure_JA_RMP83}
A.~Jayaraman, \emph{Rev. Mod. Phys.} \textbf{1983}, \emph{55}, 65.

\bibitem{highpressure_MaoHK_RMP18}
H.-K. {Mao}, X.-J. {Chen}, Y.~{Ding}, B.~{Li}, L.~{Wang}, \emph{Rev. Mod.
  Phys.} \textbf{2018}, \emph{90}, 015007.

\bibitem{H2SSC_Drozd_Nature16}
A.~P. {Drozdov}, M.~I. {Eremets}, I.~A. {Troyan}, V.~{Ksenofontov}, S.~I.
  {Shylin}, \emph{Nature} \textbf{2015}, \emph{525}, 73.

\bibitem{HPtoroomtSC_Gor_RMP18}
L.~P. {Gor'kov}, V.~Z. {Kresin}, \emph{Rev. Mod. Phys.} \textbf{2018},
  \emph{90}, 011001.

\bibitem{HgCdTe2DTI_RothA_science09}
A.~{Roth}, C.~{Br{\"u}ne}, H.~{Buhmann}, L.~W. {Molenkamp}, J.~{Maciejko},
  X.-L. {Qi}, S.-C. {Zhang}, \emph{Science} \textbf{2009}, \emph{325}, 294.

\bibitem{Sb1-xBixTI_Hsieh_nat08}
D.~{Hsieh}, D.~{Qian}, L.~{Wray}, Y.~{Xia}, Y.~S. {Hor}, R.~J. {Cava}, M.~Z.
  {Hasan}, \emph{Nature} \textbf{2008}, \emph{452}, 970.

\bibitem{Bi2Te3TI_ChenYL_science09}
Y.~L. {Chen}, J.~G. {Analytis}, J.-H. {Chu}, Z.~K. {Liu}, S.-K. {Mo}, X.~L.
  {Qi}, H.~J. {Zhang}, D.~H. {Lu}, X.~{Dai}, Z.~{Fang}, S.~C. {Zhang}, I.~R.
  {Fisher}, Z.~{Hussain}, Z.-X. {Shen}, \emph{Science} \textbf{2009},
  \emph{325}, 178.

\bibitem{Bi2Se3_XiaY_NatP09}
Y.~{Xia}, D.~{Qian}, D.~{Hsieh}, L.~{Wray}, A.~{Pal}, H.~{Lin}, A.~{Bansil},
  D.~{Grauer}, Y.~S. {Hor}, R.~J. {Cava}, M.~Z. {Hasan}, \emph{Nat. Phys.}
  \textbf{2009}, \emph{5}, 398.

\bibitem{Sb2Te3_HsiehD_PRL09}
D.~Hsieh, Y.~Xia, D.~Qian, L.~Wray, F.~Meier, J.~H. Dil, J.~Osterwalder,
  L.~Patthey, A.~V. Fedorov, H.~Lin, A.~Bansil, D.~Grauer, Y.~S. Hor, R.~J.
  Cava, M.~Z. Hasan, \emph{Phys. Rev. Lett.} \textbf{2009}, \emph{103}, 146401.

\bibitem{TI_Hasan_RMP10}
M.~Z. Hasan, C.~L. Kane, \emph{Rev. Mod. Phys.} \textbf{2010}, \emph{82}, 3045.

\bibitem{Bi2Te3SC_ZhangJL_PNAS11}
J.~L. {Zhang}, S.~J. {Zhang}, H.~M. {Weng}, W.~{Zhang}, L.~X. {Yang}, Q.~Q.
  {Liu}, S.~M. {Feng}, X.~C. {Wang}, R.~C. {Yu}, L.~Z. {Cao}, L.~{Wang}, W.~G.
  {Yang}, H.~Z. {Liu}, W.~Y. {Zhao}, S.~C. {Zhang}, X.~{Dai}, Z.~{Fang}, C.~Q.
  {Jin}, \emph{Proc. Natl. Acad. Sci. U.S.A.} \textbf{2011}, \emph{108}, 24.

\bibitem{Bi2Te3Hall_ZhangC_PRB11}
C.~{Zhang}, L.~{Sun}, Z.~{Chen}, X.~{Zhou}, Q.~{Wu}, W.~{Yi}, J.~{Guo},
  X.~{Dong}, Z.~{Zhao}, \emph{Phys. Rev. B} \textbf{2011}, \emph{83}, 140504.

\bibitem{Bi2Te3_MK_PRB14}
K.~Matsubayashi, T.~Terai, J.~S. Zhou, Y.~Uwatoko, \emph{Phys. Rev. B}
  \textbf{2014}, \emph{90}, 125126.

\bibitem{Bi2Se3SC_Kevin_PRL13}
K.~Kirshenbaum, P.~S. Syers, A.~P. Hope, N.~P. Butch, J.~R. Jeffries, S.~T.
  Weir, J.~J. Hamlin, M.~B. Maple, Y.~K. Vohra, J.~Paglione, \emph{Phys. Rev.
  Lett.} \textbf{2013}, \emph{111}, 087001.

\bibitem{Sb2Te3_ZhuJ_ScientificR}
J.~Zhu, J.~L. Zhang, P.~P. Kong, S.~J. Zhang, X.~H. Yu, J.~L. Zhu, Q.~Q. Liu,
  X.~Li, R.~C. Yu, R.~Ahuja, W.~G. Yang, G.~Y. Shen, H.~K. Mao, H.~M. Weng,
  X.~Dai, Z.~Fang, Y.~S. Zhao, C.~Q. Jin, \emph{Sci. Rep.} \textbf{2013},
  \emph{3}, 2016.

\bibitem{Sb2Te3P_ZhaoJG_IC11}
J.~Zhao, H.~Liu, L.~Ehm, Z.~Chen, S.~Sinogeikin, Y.~Zhao, G.~Gu, \emph{Inorg.
  Chem.} \textbf{2011}, \emph{50}, 11291.

\bibitem{BiTeISC_QiYP_AM17}
Y.~Qi, W.~Shi, P.~G. Naumov, N.~Kumar, R.~Sankar, W.~Schnelle, C.~Shekhar,
  F.-C. Chou, C.~Felser, B.~Yan, S.~A. Medvedev, \emph{Adv. Mater.}
  \textbf{2017}, \emph{29}, 1605965.

\bibitem{BiTeClSC_YangJJ_PRB16}
J.-J. {Ying}, V.~V. {Struzhkin}, Z.-Y. {Cao}, A.~F. {Goncharov}, H.-K. {Mao},
  F.~{Chen}, X.-H. {Chen}, A.~G. {Gavriliuk}, X.-J. {Chen}, \emph{Phys. Rev. B}
  \textbf{2016}, \emph{93}, 100504.

\bibitem{Cd3As2_HeLP2016QuantMater}
L.~P. He, Y.~T. Jia, S.~J. Zhang, X.~C. Hong, C.~Q. Jin, S.~Y. Li, \emph{npj
  {Q}uantum {M}aterials} \textbf{2016}, \emph{1}, 16014.

\bibitem{ZrTe5HP_zhouYH_PANS}
Y.~{Zhou}, J.~{Wu}, W.~{Ning}, N.~{Li}, Y.~{Du}, X.~{Chen}, R.~{Zhang},
  Z.~{Chi}, X.~{Wang}, X.~{Zhu}, P.~{Lu}, C.~{Ji}, X.~{Wan}, Z.~{Yang},
  J.~{Sun}, W.~{Yang}, M.~{Tian}, Y.~{Zhang}, H.-k. {Mao}, \emph{Proc. Natl.
  Acad. Sci. U.S.A.} \textbf{2016}, \emph{113}, 2904.

\bibitem{HfTe5SC_QiY_PRB16}
Y.~{Qi}, W.~{Shi}, P.~G. {Naumov}, N.~{Kumar}, W.~{Schnelle}, O.~{Barkalov},
  C.~{Shekhar}, H.~{Borrmann}, C.~{Felser}, B.~{Yan}, S.~A. {Medvedev},
  \emph{Phys. Rev. B} \textbf{2016}, \emph{94}, 054517.

\bibitem{TaPSC_LiY_npjQM17}
Y.~{Li}, Y.~{Zhou}, Z.~{Guo}, F.~{Han}, X.~{Chen}, P.~{Lu}, X.~{Wang}, C.~{An},
  Y.~{Zhou}, J.~{Xing}, G.~{Du}, X.~{Zhu}, H.~{Yang}, J.~{Sun}, Z.~{Yang},
  W.~{Yang}, H.-K. {Mao}, Y.~{Zhang}, H.-H. {Wen}, \emph{npj {Q}uantum
  {M}aterials} \textbf{2017}, \emph{2}, 66.

\bibitem{MoTe2_Qiy_NC16}
Y.~{Qi}, P.~G. {Naumov}, M.~N. {Ali}, C.~R. {Rajamathi}, W.~{Schnelle},
  O.~{Barkalov}, M.~{Hanfland}, S.-C. {Wu}, C.~{Shekhar}, Y.~{Sun},
  V.~{S{\"u}{\ss}}, M.~{Schmidt}, U.~{Schwarz}, E.~{Pippel}, P.~{Werner},
  R.~{Hillebrand}, T.~{F{\"o}rster}, E.~{Kampert}, S.~{Parkin}, R.~J. {Cava},
  C.~{Felser}, B.~{Yan}, S.~A. {Medvedev}, \emph{Nat. Commun.} \textbf{2016},
  \emph{7}, 11038.

\bibitem{LaBiSC_TaftiFF_PRB17}
F.~F. Tafti, M.~S. Torikachvili, R.~L. Stillwell, B.~Baer, E.~Stavrou, S.~T.
  Weir, Y.~K. Vohra, H.-Y. Yang, E.~F. McDonnell, S.~K. Kushwaha, Q.~D. Gibson,
  R.~J. Cava, J.~R. Jeffries, \emph{Phys. Rev. B} \textbf{2017}, \emph{95},
  014507.

\bibitem{WTe2_DFKang_NatC}
D.~Kang, Y.~Zhou, W.~Yi, C.~Yang, J.~Guo, Y.~Shi, S.~Zhang, Z.~Wang, C.~Zhang,
  S.~Jiang, A.~Li, K.~Yang, Q.~Wu, G.~Zhang, L.~L. Sun, Z.~X. Zhao, \emph{Nat.
  Commun.} \textbf{2015}, \emph{6}, 7804.

\bibitem{WTe2_XCPan_NatC}
X.~C. Pan, X.~L. Chen, H.~M. Liu, Y.~Q. Feng, Z.~X. Wei, Y.~H. Zhou, Z.~H. Chi,
  L.~Pi, F.~Yen, F.~Q. Song, X.~G. Wan, Z.~R. Yang, B.~G. Wang, G.~H. Wang,
  Y.~H. Zhang, \emph{Nat. Commun.} \textbf{2015}, \emph{6}, 7805.

\bibitem{MoPHPSC_ChiZ_npj18}
Z.~{Chi}, X.~{Chen}, C.~{An}, L.~{Yang}, J.~{Zhao}, Z.~{Feng}, Y.~{Zhou},
  Y.~{Zhou}, C.~{Gu}, B.~{Zhang}, Y.~{Yuan}, C.~{Kenney-Benson}, W.~{Yang},
  G.~{Wu}, X.~{Wan}, Y.~{Shi}, X.~{Yang}, Z.~{Yang}, \emph{npj {Q}uantum
  {M}aterials} \textbf{2018}, \emph{3}, 28.

\bibitem{NbAs2SC_Liyp}
Y.~{Li}, C.~{An}, C.~{Hua}, X.~{Chen}, Y.~{Zhou}, Y.~{Zhou}, R.~{Zhang},
  C.~{Park}, Z.~{Wang}, Y.~{Lu}, Y.~{Zheng}, Z.~{Yang}, Z.-A. {Xu}, \emph{npj
  {Q}uantum {M}aterials} \textbf{2018}, \emph{3}, 58.

\bibitem{CuxBi2Se3_HorYS_PRL10}
Y.~S. Hor, A.~J. Williams, J.~G. Checkelsky, P.~Roushan, J.~Seo, Q.~Xu, H.~W.
  Zandbergen, A.~Yazdani, N.~P. Ong, R.~J. Cava, \emph{Phys. Rev. Lett.}
  \textbf{2010}, \emph{104}, 057001.

\bibitem{SrxBi2Se3_Shruti_PRB15}
Shruti, V.~K. Maurya, P.~Neha, P.~Srivastava, S.~Patnaik, \emph{Phys. Rev. B}
  \textbf{2015}, \emph{92}, 020506.

\bibitem{Bi2Se3struc_Vilaplana_PRB11}
R.~{Vilaplana}, D.~{Santamar{\'{\i}}a-P{\'e}rez}, O.~{Gomis}, F.~J.
  {Manj{\'o}n}, J.~{Gonz{\'a}lez}, A.~{Segura}, A.~{Mu{\~n}oz},
  P.~{Rodr{\'{\i}}guez-Hern{\'a}ndez}, E.~{P{\'e}rez-Gonz{\'a}lez},
  V.~{Mar{\'{\i}}n-Borr{\'a}s}, V.~{Mu{\~n}oz-Sanjose}, C.~{Drasar},
  V.~{Kucek}, \emph{Phys. Rev. B} \textbf{2011}, \emph{84}, 184110.

\bibitem{BiTeIAPRES_Ishizaka_natM11}
K.~{Ishizaka}, M.~S. {Bahramy}, H.~{Murakawa}, M.~{Sakano}, T.~{Shimojima},
  T.~{Sonobe}, K.~{Koizumi}, S.~{Shin}, H.~{Miyahara}, A.~{Kimura},
  K.~{Miyamoto}, T.~{Okuda}, H.~{Namatame}, M.~{Taniguchi}, R.~{Arita},
  N.~{Nagaosa}, K.~{Kobayashi}, Y.~{Murakami}, R.~{Kumai}, Y.~{Kaneko},
  Y.~{Onose}, Y.~{Tokura}, \emph{Nat. Mater.} \textbf{2011}, \emph{10}, 521.

\bibitem{BiTeIAPRES_Sakano_PRL13}
M.~Sakano, M.~S. Bahramy, A.~Katayama, T.~Shimojima, H.~Murakawa, Y.~Kaneko,
  W.~Malaeb, S.~Shin, K.~Ono, H.~Kumigashira, R.~Arita, N.~Nagaosa, H.~Y.
  Hwang, Y.~Tokura, K.~Ishizaka, \emph{Phys. Rev. Lett.} \textbf{2013},
  \emph{110}, 107204.

\bibitem{BiTeCl_ChenYL_NatP13}
Y.~L. {Chen}, M.~{Kanou}, Z.~K. {Liu}, H.~J. {Zhang}, J.~A. {Sobota},
  D.~{Leuenberger}, S.~K. {Mo}, B.~{Zhou}, S.-L. {Yang}, P.~S. {Kirchmann},
  D.~H. {Lu}, R.~G. {Moore}, Z.~{Hussain}, Z.~X. {Shen}, X.~L. {Qi},
  T.~{Sasagawa}, \emph{Nat. Phys.} \textbf{2013}, \emph{9}, 704.

\bibitem{Na3Bi_LiuZK_science14}
Z.~K. Liu, B.~B. Zhou, Y.~Zhang, Z.~J. Wang, H.~Z. Weng, D.~Prabhakaran, S.~K.
  Mo, Z.~X. Shen, Z.~Fang, X.~Dai, Z.~Hussain, Y.~L. Chen, \emph{Science}
  \textbf{2014}, \emph{343}, 864.

\bibitem{Cd3As2_LiuZK_NatM14}
Z.~K. {Liu}, J.~{Jiang}, B.~{Zhou}, Z.~J. {Wang}, Y.~{Zhang}, H.~M. {Weng},
  D.~{Prabhakaran}, S.-K. {Mo}, H.~{Peng}, P.~{Dudin}, T.~{Kim}, M.~{Hoesch},
  Z.~{Fang}, X.~{Dai}, Z.~X. {Shen}, D.~L. {Feng}, Z.~{Hussain}, Y.~L. {Chen},
  \emph{Nat. Mater.} \textbf{2014}, \emph{13}, 677.

\bibitem{ZrTe5ARPES_LiQ_NP16}
Q.~{Li}, D.~E. {Kharzeev}, C.~{Zhang}, Y.~{Huang}, I.~{Pletikosi{\'c}}, A.~V.
  {Fedorov}, R.~D. {Zhong}, J.~A. {Schneeloch}, G.~D. {Gu}, T.~{Valla},
  \emph{Nat. Phys.} \textbf{2016}, \emph{12}, 550.

\bibitem{WSMWanXG_PRB}
A.~V. X.~G.~Wan, A. M.~Turner, S.~Y. Savrasov, \emph{Phys. Rev. B}
  \textbf{2011}, \emph{83}, 205101.

\bibitem{WeylSemi_WHM_PRX}
H.~Weng, C.~Fang, Z.~Fang, B.~A. Bernevig, X.~Dai, \emph{Phys. Rev. X}
  \textbf{2015}, \emph{5}, 011029.

\bibitem{TaAs_Hasan_Science}
S.-Y. {Xu}, I.~{Belopolski}, N.~{Alidoust}, M.~{Neupane}, G.~{Bian},
  C.~{Zhang}, R.~{Sankar}, G.~{Chang}, Z.~{Yuan}, C.-C. {Lee}, S.-M. {Huang},
  H.~{Zheng}, J.~{Ma}, D.~S. {Sanchez}, B.~{Wang}, A.~{Bansil}, F.~{Chou},
  P.~P. {Shibayev}, H.~{Lin}, S.~{Jia}, M.~Z. {Hasan}, \emph{Science}
  \textbf{2015}, \emph{349}, 613.

\bibitem{TaP_XuSY_SciAd15}
S.-Y. {Xu}, I.~{Belopolski}, D.~S. {Sanchez}, C.~{Zhang}, G.~{Chang}, C.~{Guo},
  G.~{Bian}, Z.~{Yuan}, H.~{Lu}, T.-R. {Chang}, P.~P. {Shibayev}, M.~L.
  {Prokopovych}, N.~{Alidoust}, H.~{Zheng}, C.-C. {Lee}, S.-M. {Huang},
  R.~{Sankar}, F.~{Chou}, C.-H. {Hsu}, H.-T. {Jeng}, A.~{Bansil}, T.~{Neupert},
  V.~N. {Strocov}, H.~{Lin}, S.~{Jia}, M.~Z. {Hasan}, \emph{Sci. Adv.}
  \textbf{2015}, \emph{1}, e1501092.

\bibitem{NbAs_XuSY_NatP15}
S.-Y. {Xu}, N.~{Alidoust}, I.~{Belopolski}, Z.~{Yuan}, G.~{Bian}, T.-R.
  {Chang}, H.~{Zheng}, V.~N. {Strocov}, D.~S. {Sanchez}, G.~{Chang},
  C.~{Zhang}, D.~{Mou}, Y.~{Wu}, L.~{Huang}, C.-C. {Lee}, S.-M. {Huang},
  B.~{Wang}, A.~{Bansil}, H.-T. {Jeng}, T.~{Neupert}, A.~{Kaminski}, H.~{Lin},
  S.~{Jia}, M.~{Zahid Hasan}, \emph{Nat. Phys.} \textbf{2015}, \emph{11}, 748.

\bibitem{NbPFA_XDF_CPL15}
D.-F. {Xu}, Y.-P. {Du}, Z.~{Wang}, Y.-P. {Li}, X.-H. {Niu}, Q.~{Yao},
  P.~{Dudin}, Z.-A. {Xu}, X.-G. {Wan}, D.-L. {Feng}, \emph{Chin. Phys. Lett.}
  \textbf{2015}, \emph{32}, 107101.

\bibitem{NbP_WZ_PRB}
Z.~{Wang}, Y.~{Zheng}, Z.~{Shen}, Y.~{Lu}, H.~{Fang}, F.~{Sheng}, Y.~{Zhou},
  X.~{Yang}, Y.~{Li}, C.~{Feng}, Z.-A. {Xu}, \emph{Phys. Rev. B} \textbf{2016},
  \emph{93}, 121112.

\bibitem{currentjeting_HuJ_PRL05}
J.~{Hu}, T.~F. {Rosenbaum}, J.~B. {Betts}, \emph{Phys. Rev. Lett.}
  \textbf{2005}, \emph{95}, 186603.

\bibitem{NMRinWSMs_dRD_NJP16}
R.~D. {dos Reis}, M.~O. {Ajeesh}, N.~{Kumar}, F.~{Arnold}, C.~{Shekhar},
  M.~{Naumann}, M.~{Schmidt}, M.~{Nicklas}, E.~{Hassinger}, \emph{New J. Phys.}
  \textbf{2016}, \emph{18}, 085006.

\bibitem{NMR_YPL_FP17}
Y.~{Li}, Z.~{Wang}, P.~{Li}, X.~{Yang}, Z.~{Shen}, F.~{Sheng}, X.~{Li},
  Y.~{Lu}, Y.~{Zheng}, Z.-A. {Xu}, \emph{Front. Phys.} \textbf{2017},
  \emph{12}, 127205.

\bibitem{Cd3As2break_ZhangS_PRB15}
S.~{Zhang}, Q.~{Wu}, L.~{Schoop}, M.~N. {Ali}, Y.~{Shi}, N.~{Ni}, Q.~{Gibson},
  S.~{Jiang}, V.~{Sidorov}, W.~{Yi}, J.~{Guo}, Y.~{Zhou}, D.~{Wu}, P.~{Gao},
  D.~{Gu}, C.~{Zhang}, S.~{Jiang}, K.~{Yang}, A.~{Li}, Y.~{Li}, X.~{Li},
  J.~{Liu}, X.~{Dai}, Z.~{Fang}, R.~J. {Cava}, L.~{Sun}, Z.~{Zhao}, \emph{Phys.
  Rev. B} \textbf{2015}, \emph{91}, 165133.

\bibitem{Cd3As2TSCTheory_Kob_PRL15}
S.~Kobayashi, M.~Sato, \emph{Phys. Rev. Lett.} \textbf{2015}, \emph{115},
  187001.

\bibitem{WeylSCMF_SauJ_PRB12}
J.~D. Sau, S.~Tewari, \emph{Phys. Rev. B} \textbf{2012}, \emph{86}, 104509.

\bibitem{proximityEfWeylSC_ChenA_PRB16}
A.~Chen, M.~Franz, \emph{Phys. Rev. B} \textbf{2016}, \emph{93}, 201105.

\bibitem{NbAs_ZhangJ_CPL15}
J.~{Zhang}, F.-L. {Liu}, J.-K. {Dong}, Y.~{Xu}, N.-N. {Li}, W.-G. {Yang}, S.-Y.
  {Li}, \emph{Chin. Phys. Lett.} \textbf{2015}, \emph{32}, 097102.

\bibitem{WTe2WSM_WuL_PRB16}
Y.~{Wu}, D.~{Mou}, N.~H. {Jo}, K.~{Sun}, L.~{Huang}, S.~L. {Bud'ko}, P.~C.
  {Canfield}, A.~{Kaminski}, \emph{Phys. Rev. B} \textbf{2016}, \emph{94},
  121113.

\bibitem{MoTe2WSM_Dengk_NP16}
K.~{Deng}, G.~{Wan}, P.~{Deng}, K.~{Zhang}, S.~{Ding}, E.~{Wang}, M.~{Yan},
  H.~{Huang}, H.~{Zhang}, Z.~{Xu}, J.~{Denlinger}, A.~{Fedorov}, H.~{Yang},
  W.~{Duan}, H.~{Yao}, Y.~{Wu}, S.~{Fan}, H.~{Zhang}, X.~{Chen}, S.~{Zhou},
  \emph{Nat. Phys.} \textbf{2016}, \emph{12}, 1105.

\bibitem{MoxWTe2SM_ChangTR_nc16}
T.-R. {Chang}, S.-Y. {Xu}, G.~{Chang}, C.-C. {Lee}, S.-M. {Huang}, B.~{Wang},
  G.~{Bian}, H.~{Zheng}, D.~S. {Sanchez}, I.~{Belopolski}, N.~{Alidoust},
  M.~{Neupane}, A.~{Bansil}, H.-T. {Jeng}, H.~{Lin}, M.~{Zahid Hasan},
  \emph{Nat. Commun.} \textbf{2016}, \emph{7}, 10639.

\bibitem{Ta3S2TpeIIWS_ChangG_SciAdv16}
G.~{Chang}, S.-Y. {Xu}, D.~S. {Sanchez}, S.-M. {Huang}, C.-C. {Lee}, T.-R.
  {Chang}, G.~{Bian}, H.~{Zheng}, I.~{Belopolski}, N.~{Alidoust}, H.-T. {Jeng},
  A.~{Bansil}, H.~{Lin}, M.~Z. {Hasan}, \emph{Sci. Adv.} \textbf{2016},
  \emph{2}, e1600295.

\bibitem{typeIIWeylS_SoluAA_nat15}
A.~A. {Soluyanov}, D.~{Gresch}, Z.~{Wang}, Q.~{Wu}, M.~{Troyer}, X.~{Dai},
  B.~A. {Bernevig}, \emph{Nature} \textbf{2015}, \emph{527}, 495.

\bibitem{WTe2_Ali_Nat14}
M.~N. {Ali}, J.~{Xiong}, S.~{Flynn}, J.~{Tao}, Q.~D. {Gibson}, L.~M. {Schoop},
  T.~{Liang}, N.~{Haldolaarachchige}, M.~{Hirschberger}, N.~P. {Ong}, R.~J.
  {Cava}, \emph{Nature} \textbf{2014}, \emph{514}, 205.

\bibitem{MoTe2ARPES_DengK_NP16}
K.~{Deng}, G.~{Wan}, P.~{Deng}, K.~{Zhang}, S.~{Ding}, E.~{Wang}, M.~{Yan},
  H.~{Huang}, H.~{Zhang}, Z.~{Xu}, J.~{Denlinger}, A.~{Fedorov}, H.~{Yang},
  W.~{Duan}, H.~{Yao}, Y.~{Wu}, S.~{Fan}, H.~{Zhang}, X.~{Chen}, S.~{Zhou},
  \emph{Nat. Phys.} \textbf{2016}, \emph{12}, 1105.

\bibitem{TdMoTe2ARPES_HuangL_NM16}
L.~{Huang}, T.~M. {McCormick}, M.~{Ochi}, Z.~{Zhao}, M.-T. {Suzuki},
  R.~{Arita}, Y.~{Wu}, D.~{Mou}, H.~{Cao}, J.~{Yan}, N.~{Trivedi},
  A.~{Kaminski}, \emph{Nat. Mater.} \textbf{2016}, \emph{15}, 1155.

\bibitem{LaBiARPES_NayakJ_ncom17}
J.~{Nayak}, S.-C. {Wu}, N.~{Kumar}, C.~{Shekhar}, S.~{Singh}, J.~{Fink},
  E.~E.~D. {Rienks}, G.~H. {Fecher}, S.~S.~P. {Parkin}, B.~{Yan}, C.~{Felser},
  \emph{Nat. Commun.} \textbf{2017}, \emph{8}, 13942.

\bibitem{triply_BQLv_nature17}
B.~Q. {Lv}, Z.-L. {Feng}, Q.-N. {Xu}, X.~{Gao}, J.-Z. {Ma}, L.-Y. {Kong},
  P.~{Richard}, Y.-B. {Huang}, V.~N. {Strocov}, C.~{Fang}, H.-M. {Weng}, Y.-G.
  {Shi}, T.~{Qian}, H.~{Ding}, \emph{Nature} \textbf{2017}, \emph{546}, 627.

\bibitem{NbAs2DiracNL_YMShao_PNAS19}
Y.~{Shao}, Z.~{Sun}, Y.~{Wang}, C.~{Xu}, R.~{Sankar}, A.~J. {Breindel},
  C.~{Cao}, M.~M. {Fogler}, A.~J. {Millis}, F.~{Chou}, Z.~{Li}, T.~{Timusk},
  M.~B. {Maple}, D.~N. {Basov}, \emph{Proc. Natl. Acad. Sci. U.S.A.}
  \textbf{2019}, \emph{116}, 1168.

\bibitem{NbAs2family_YupengLi_arXiv}
Y.~P. Li, Z.~Wang, Y.~H. Lu, X.~J. Yang, Z.~X. Shen, F.~Sheng, C.~M. Feng,
  Y.~Zheng, Z.~A. Xu, \emph{arXiv:1603.04056} .

\bibitem{NbAs2_ShenB_PRB}
B.~{Shen}, X.~{Deng}, G.~{Kotliar}, N.~{Ni}, \emph{Phys. Rev. B} \textbf{2016},
  \emph{93}, 195119.

\bibitem{NbAs2Weyl_GreschD_NJP17}
D.~{Gresch}, Q.~{Wu}, G.~W. {Winkler}, A.~A. {Soluyanov}, \emph{New. J. Phys.}
  \textbf{2017}, \emph{19}, 035001.

\bibitem{TipSC_WangH_SB18}
H.~{Wang}, L.~{Ma}, J.~{Wang}, \emph{Sci. Bull.} \textbf{2018}, \emph{63}, 1141
  .

\bibitem{Cd3As2tipSC_WangH_NM16}
H.~{Wang}, H.~{Wang}, H.~{Liu}, H.~{Lu}, W.~{Yang}, S.~{Jia}, X.-J. {Liu},
  X.~C. {Xie}, J.~{Wei}, J.~{Wang}, \emph{Nat. Mater.} \textbf{2016},
  \emph{15}, 38.

\bibitem{Cd3As2tipSC_Aggar_NM16}
L.~{Aggarwal}, A.~{Gaurav}, G.~S. {Thakur}, Z.~{Haque}, A.~K. {Ganguli},
  G.~{Sheet}, \emph{Nat. Mater.} \textbf{2016}, \emph{15}, 32.

\bibitem{LaOFeAsF_Kamihara_JACS08}
Y.~Kamihara, T.~Watanabe, M.~Hirano, H.~Hosono, \emph{J. Am. Chem. Soc.}
  \textbf{2008}, \emph{130}, 3296.

\bibitem{PdopingSC_JiangS_JPCM09}
S.~{Jiang}, H.~{Xing}, G.~{Xuan}, C.~{Wang}, Z.~{Ren}, C.~{Feng}, J.~{Dai},
  Z.~{Xu}, G.~{Cao}, \emph{J. Phys.: Condens. Matter.} \textbf{2009},
  \emph{21}, 382203.

\bibitem{QAHinCrxBi2Se3_ChangCZ_Science13}
C.-Z. {Chang}, J.~{Zhang}, X.~{Feng}, J.~{Shen}, Z.~{Zhang}, M.~{Guo}, K.~{Li},
  Y.~{Ou}, P.~{Wei}, L.-L. {Wang}, Z.-Q. {Ji}, Y.~{Feng}, S.~{Ji}, X.~{Chen},
  J.~{Jia}, X.~{Dai}, Z.~{Fang}, S.-C. {Zhang}, K.~{He}, Y.~{Wang}, L.~{Lu},
  X.-C. {Ma}, Q.-K. {Xue}, \emph{Science} \textbf{2013}, \emph{340}, 167.

\bibitem{DMS_DietlT_NatM10}
T.~{Dietl}, \emph{Nat. Mater.} \textbf{2010}, \emph{9}, 965.

\bibitem{SCdopedTM_Sasaki_PhysC15}
S.~{Sasaki}, T.~{Mizushima}, \emph{Physica C} \textbf{2015}, \emph{514}, 206.

\bibitem{CuxBi2Se3ECQ_Kriener_PRB11}
M.~{Kriener}, K.~{Segawa}, Z.~{Ren}, S.~{Sasaki}, S.~{Wada}, S.~{Kuwabata},
  Y.~{Ando}, \emph{Phys. Rev. B} \textbf{2011}, \emph{84}, 054513.

\bibitem{SrxBi2Se3SC_Liuzh_JACS15}
Z.~Liu, X.~Yao, J.~Shao, M.~Zuo, L.~Pi, S.~Tan, C.~Zhang, Y.~Zhang, \emph{J.
  Am. Chem. Soc.} \textbf{2015}, \emph{137}, 10512.

\bibitem{Nb0.25Bi2Se3_QiuYS_arXiv15}
Y.~{Qiu}, K.~{Nocona Sanders}, J.~{Dai}, J.~E. {Medvedeva}, W.~{Wu},
  P.~{Ghaemi}, T.~{Vojta}, Y.~{San Hor}, \emph{arXiv:1512.03519} .

\bibitem{Tl0.6Bi2Te3SC_WangZW_CM16}
Z.~Wang, A.~A. Taskin, T.~Fr\"{o}lich, M.~Braden, Y.~Ando, \emph{Chem. Mater.}
  \textbf{2016}, \emph{28}, 779.

\bibitem{Tl0.5Bi2Te3ARPES_Trang_prb16}
C.~X. Trang, Z.~Wang, D.~Takane, K.~Nakayama, S.~Souma, T.~Sato, T.~Takahashi,
  A.~A. Taskin, Y.~Ando, \emph{Phys. Rev. B} \textbf{2016}, \emph{93}, 241103.

\bibitem{Cux(PbSe)5(Bi2Se3)6_Sasaki_PRB14}
S.~Sasaki, K.~Segawa, Y.~Ando, \emph{Phys. Rev. B} \textbf{2014}, \emph{90},
  220504.

\bibitem{Sn1-xInxTeSC_Ehhanced_PRB09}
A.~S. Erickson, J.-H. Chu, M.~F. Toney, T.~H. Geballe, I.~R. Fisher,
  \emph{Phys. Rev. B} \textbf{2009}, \emph{79}, 024520.

\bibitem{Sn0.6In0.4Te_Balak_PRB13}
G.~Balakrishnan, L.~Bawden, S.~Cavendish, M.~R. Lees, \emph{Phys. Rev. B}
  \textbf{2013}, \emph{87}, 140507.

\bibitem{(PbSn)1-xInxTeSC_ZhangRD_PRB14}
R.~D. Zhong, J.~A. Schneeloch, T.~S. Liu, F.~E. Camino, J.~M. Tranquada, G.~D.
  Gu, \emph{Phys. Rev. B} \textbf{2014}, \emph{90}, 020505.

\bibitem{CuxBi2Se3fullgap_Kriener_PRL11}
M.~{Kriener}, K.~{Segawa}, Z.~{Ren}, S.~{Sasaki}, Y.~{Ando}, \emph{Phys. Rev.
  Lett.} \textbf{2011}, \emph{106}, 127004.

\bibitem{CuxBi2Se3sWave_LevyN_PRL13}
N.~{Levy}, T.~{Zhang}, J.~{Ha}, F.~{Sharifi}, A.~A. {Talin}, Y.~{Kuk}, J.~A.
  {Stroscio}, \emph{Phys. Rev. Lett.} \textbf{2013}, \emph{110}, 117001.

\bibitem{CuxBi2Se3SS_LahoudE_PRB13}
E.~Lahoud, E.~Maniv, M.~S. Petrushevsky, M.~Naamneh, A.~Ribak, S.~Wiedmann,
  L.~Petaccia, Z.~Salman, K.~B. Chashka, Y.~Dagan, A.~Kanigel, \emph{Phys. Rev.
  B} \textbf{2013}, \emph{88}, 195107.

\bibitem{CuxBi2Se3HPSC_BayTV_PRL12}
T.~V. Bay, T.~Naka, Y.~K. Huang, H.~Luigjes, M.~S. Golden, A.~de~Visser,
  \emph{Phys. Rev. Lett.} \textbf{2012}, \emph{108}, 057001.

\bibitem{CuxBi2Se3ZBCP_Sasaki_PRL11}
S.~{Sasaki}, M.~{Kriener}, K.~{Segawa}, K.~{Yada}, Y.~{Tanaka}, M.~{Sato},
  Y.~{Ando}, \emph{Phys. Rev. Lett.} \textbf{2011}, \emph{107}, 217001.

\bibitem{CuxBi2Se3NMR_MatanoK_NP16}
K.~{Matano}, M.~{Kriener}, K.~{Segawa}, Y.~{Ando}, G.-Q. {Zheng}, \emph{Nat.
  Phys.} \textbf{2016}, \emph{12}, 852.

\bibitem{CuxBi2Se3Thermo_YonezawaS_NP17}
S.~{Yonezawa}, K.~{Tajiri}, S.~{Nakata}, Y.~{Nagai}, Z.~{Wang}, K.~{Segawa},
  Y.~{Ando}, Y.~{Maeno}, \emph{Nat. Phys.} \textbf{2017}, \emph{13}, 123.

\bibitem{CuxBi2Se3_TaoR_PRX18}
R.~{Tao}, Y.-J. {Yan}, X.~{Liu}, Z.-W. {Wang}, Y.~{Ando}, Q.-H. {Wang},
  T.~{Zhang}, D.-L. {Feng}, \emph{Phys. Rev. X} \textbf{2018}, \emph{8},
  041024.

\bibitem{CuxBi2Se3_FuL_PRB}
L.~Fu, \emph{Phys. Rev. B} \textbf{2014}, \emph{90}, 100509.

\bibitem{(PbSe)5(Bi2Se3)6_Naka_PRL12}
K.~Nakayama, K.~Eto, Y.~Tanaka, T.~Sato, S.~Souma, T.~Takahashi, K.~Segawa,
  Y.~Ando, \emph{Phys. Rev. Lett.} \textbf{2012}, \emph{109}, 236804.

\bibitem{Cux(PbSe)5(Bi2Se3)6ARPES_Nakay_PRB15}
K.~Nakayama, H.~Kimizuka, Y.~Tanaka, T.~Sato, S.~Souma, T.~Takahashi,
  S.~Sasaki, K.~Segawa, Y.~Ando, \emph{Phys. Rev. B} \textbf{2015}, \emph{92},
  100508.

\bibitem{SrxBi2Se3NSC_PanY_SR16}
Y.~{Pan}, A.~M. {Nikitin}, G.~K. {Araizi}, Y.~K. {Huang}, Y.~{Matsushita},
  T.~{Naka}, A.~{de Visser}, \emph{Sci. Rep.} \textbf{2016}, \emph{6}, 28632.

\bibitem{SrxBi2Se3NSC_DuG_SCP17}
G.~{Du}, Y.~{Li}, J.~{Schneeloch}, R.~D. {Zhong}, G.~{Gu}, H.~{Yang}, H.~{Lin},
  H.-H. {Wen}, \emph{Sci. China Phys. Mech. Astron.} \textbf{2017}, \emph{60},
  37411.

\bibitem{NbxBi2Se3NSC_AsabaT_PRX17}
T.~{Asaba}, B.~J. {Lawson}, C.~{Tinsman}, L.~{Chen}, P.~{Corbae}, G.~{Li},
  Y.~{Qiu}, Y.~S. {Hor}, L.~{Fu}, L.~{Li}, \emph{Phys. Rev. X} \textbf{2017},
  \emph{7}, 011009.

\bibitem{Nb0.25Bi2Se3_ShenJ_npjQM17}
J.~{Shen}, W.-Y. {He}, N.~F.~Q. {Yuan}, Z.~{Huang}, C.-w. {Cho}, S.~H. {Lee},
  Y.~S. {Hor}, K.~T. {Law}, R.~{Lortz}, \emph{npj Quantum Materials}
  \textbf{2017}, \emph{2}, 59.

\bibitem{Cux(PbSe)5(Bi2Se3)6Nematic_Andersen_PRB18}
L.~{Andersen}, Z.~{Wang}, T.~{Lorenz}, Y.~{Ando}, \emph{Phys. Rev. B}
  \textbf{2018}, \emph{98}, 220512.

\bibitem{CuxBi2Se3_FuL_PRL}
L.~Fu, E.~Berg, \emph{Phys. Rev. Lett.} \textbf{2010}, \emph{105}, 097001.

\bibitem{turning_wanxg_NC}
X.~G. Wan, S.~Y. Savrasov, \emph{Nat. Commun.} \textbf{2014}, \emph{5}, 4144.

\bibitem{TCI_Ful_PRL11}
L.~Fu, \emph{Phys. Rev. Lett.} \textbf{2011}, \emph{106}, 106802.

\bibitem{SnTeTheory_Hsieh_NC12}
T.~H. {Hsieh}, H.~{Lin}, J.~{Liu}, W.~{Duan}, A.~{Bansil}, L.~{Fu}, \emph{Nat.
  Commun.} \textbf{2012}, \emph{3}, 982.

\bibitem{SnTeARPES_Tanaka_NP12}
Y.~{Tanaka}, Z.~{Ren}, T.~{Sato}, K.~{Nakayama}, S.~{Souma}, T.~{Takahashi},
  K.~{Segawa}, Y.~{Ando}, \emph{Nat. Phys.} \textbf{2012}, \emph{8}, 800.

\bibitem{Sn1-xInxTethermocond_HeLP_PRB13}
L.~P. He, Z.~Zhang, J.~Pan, X.~C. Hong, S.~Y. Zhou, S.~Y. Li, \emph{Phys. Rev.
  B} \textbf{2013}, \emph{88}, 014523.

\bibitem{Sn1-xInxTe_Novak_PRB13}
M.~{Novak}, S.~{Sasaki}, M.~{Kriener}, K.~{Segawa}, Y.~{Ando}, \emph{Phys. Rev.
  B} \textbf{2013}, \emph{88}, 140502.

\bibitem{Sn1-xInxTeUSR_Saghir_PRB14}
M.~Saghir, J.~A.~T. Barker, G.~Balakrishnan, A.~D. Hillier, M.~R. Lees,
  \emph{Phys. Rev. B} \textbf{2014}, \emph{90}, 064508.

\bibitem{Sn1-xInxTeTSC_Sasaki_PRL12}
S.~Sasaki, Z.~Ren, A.~A. Taskin, K.~Segawa, L.~Fu, Y.~Ando, \emph{Phys. Rev.
  Lett.} \textbf{2012}, \emph{109}, 217004.

\bibitem{Sn1-xInxTeARPED_Sato_PRL13}
T.~Sato, Y.~Tanaka, K.~Nakayama, S.~Souma, T.~Takahashi, S.~Sasaki, Z.~Ren,
  A.~A. Taskin, K.~Segawa, Y.~Ando, \emph{Phys. Rev. Lett.} \textbf{2013},
  \emph{110}, 206804.

\bibitem{Pn1-xSnxTetheory_Tanaka_PRB13}
Y.~Tanaka, T.~Sato, K.~Nakayama, S.~Souma, T.~Takahashi, Z.~Ren, M.~Novak,
  K.~Segawa, Y.~Ando, \emph{Phys. Rev. B} \textbf{2013}, \emph{87}, 155105.

\bibitem{SnTeTCI_Tanaka_NP12}
Y.~{Tanaka}, Z.~{Ren}, T.~{Sato}, K.~{Nakayama}, S.~{Souma}, T.~{Takahashi},
  K.~{Segawa}, Y.~{Ando}, \emph{Nat. Phys.} \textbf{2012}, \emph{8}, 800.

\bibitem{Pn1-xSnxSeTheoryArpes_Dziawa_NM12}
P.~{Dziawa}, B.~J. {Kowalski}, K.~{Dybko}, R.~{Buczko}, A.~{Szczerbakow},
  M.~{Szot}, E.~{{\L}usakowska}, T.~{Balasubramanian}, B.~M. {Wojek}, M.~H.
  {Berntsen}, O.~{Tjernberg}, T.~{Story}, \emph{Nat. Mater.} \textbf{2012},
  \emph{11}, 1023.

\bibitem{Pn1-xSnxSetrans_LiangT_NC13}
T.~{Liang}, Q.~{Gibson}, J.~{Xiong}, M.~{Hirschberger}, S.~P. {Koduvayur},
  R.~J. {Cava}, N.~P. {Ong}, \emph{Nat. Commun.} \textbf{2013}, \emph{4}, 2696.

\bibitem{(PbSn)1-xInxTeFualgap_GuanD_PRB15}
G.~{Du}, Z.~{Du}, D.~{Fang}, H.~{Yang}, R.~D. {Zhong}, J.~{Schneeloch}, G.~D.
  {Gu}, H.-H. {Wen}, \emph{Phys. Rev. B} \textbf{2015}, \emph{92}, 020512.

\bibitem{Solidgate_AhnCH_RMP06}
C.~H. Ahn, A.~Bhattacharya, M.~Di~Ventra, J.~N. Eckstein, C.~D. Frisbie, M.~E.
  Gershenson, A.~M. Goldman, I.~H. Inoue, J.~Mannhart, A.~J. Millis, A.~F.
  Morpurgo, D.~Natelson, J.-M. Triscone, \emph{Rev. Mod. Phys.} \textbf{2006},
  \emph{78}, 1185.

\bibitem{2DSC_Saito_NRM17}
Y.~{Saito}, T.~{Nojima}, Y.~{Iwasa}, \emph{Nat. Rev. Mater.} \textbf{2017},
  \emph{2}, 16094.

\bibitem{GateSummary_SaitoY_SST16}
Y.~{Saito}, T.~{Nojima}, Y.~{Iwasa}, \emph{Supercond. Sci. Technol.}
  \textbf{2016}, \emph{29}, 093001.

\bibitem{FeSeSolidion_LeiB_PRB17}
B.~Lei, N.~Z. Wang, C.~Shang, F.~B. Meng, L.~K. Ma, X.~G. Luo, T.~Wu, Z.~Sun,
  Y.~Wang, Z.~Jiang, B.~H. Mao, Z.~Liu, Y.~J. Yu, Y.~B. Zhang, X.~H. Chen,
  \emph{Phys. Rev. B} \textbf{2017}, \emph{95}, 020503.

\bibitem{FeSeTeSolidion_ZhuCS_PRB17}
C.~S. Zhu, J.~H. Cui, B.~Lei, N.~Z. Wang, C.~Shang, F.~B. Meng, L.~K. Ma, X.~G.
  Luo, T.~Wu, Z.~Sun, X.~H. Chen, \emph{Phys. Rev. B} \textbf{2017}, \emph{95},
  174513.

\bibitem{FeSeSolidion_YingTP_PRL18}
T.~P. Ying, M.~X. Wang, X.~X. Wu, Z.~Y. Zhao, Z.~Z. Zhang, B.~Q. Song, Y.~C.
  Li, B.~Lei, Q.~Li, Y.~Yu, E.~J. Cheng, Z.~H. An, Y.~Zhang, X.~Y. Jia,
  W.~Yang, X.~H. Chen, S.~Y. Li, \emph{Phys. Rev. Lett.} \textbf{2018},
  \emph{121}, 207003.

\bibitem{LiFeOHFeSeSolidion_MaLK_ArXiv18}
L.~K. {Ma}, B.~{Lei}, N.~Z. {Wang}, K.~S. {Yang}, D.~Y. {Liu}, F.~B. {Meng},
  C.~{Shang}, Z.~L. {Sun}, J.~H. {Cui}, C.~S. {Zhu}, T.~{Wu}, Z.~{Sun}, L.~J.
  {Zou}, X.~H. {Chen}, \emph{arXiv:1808.06051} .

\bibitem{Ionliquidgating_Ueno_jpsj14}
K.~{Ueno}, H.~{Shimotani}, H.~{Yuan}, J.~{Ye}, M.~{Kawasaki}, Y.~{Iwasa},
  \emph{J. Phys. Soc. Jpn.} \textbf{2014}, \emph{83}, 032001.

\bibitem{Graphene_sarma_RMP11}
S.~Das~Sarma, S.~Adam, E.~H. Hwang, E.~Rossi, \emph{Rev. Mod. Phys.}
  \textbf{2011}, \emph{83}, 407.

\bibitem{Graphene_Castro_RMP09}
A.~H. Castro~Neto, F.~Guinea, N.~M.~R. Peres, K.~S. Novoselov, A.~K. Geim,
  \emph{Rev. Mod. Phys.} \textbf{2009}, \emph{81}, 109.

\bibitem{grapheneSC_CaoY_nature18}
Y.~{Cao}, V.~{Fatemi}, S.~{Fang}, K.~{Watanabe}, T.~{Taniguchi}, E.~{Kaxiras},
  P.~{Jarillo-Herrero}, \emph{Nature} \textbf{2018}, \emph{556}, 43.

\bibitem{grapheneSCMott_CaoY_Nature18}
Y.~{Cao}, V.~{Fatemi}, A.~{Demir}, S.~{Fang}, S.~L. {Tomarken}, J.~Y. {Luo},
  J.~D. {Sanchez-Yamagishi}, K.~{Watanabe}, T.~{Taniguchi}, E.~{Kaxiras}, R.~C.
  {Ashoori}, P.~{Jarillo-Herrero}, \emph{Nature} \textbf{2018}, \emph{556}, 80.

\bibitem{TSCgraphene_XuC_PRL18}
C.~Xu, L.~Balents, \emph{Phys. Rev. Lett.} \textbf{2018}, \emph{121}, 087001.

\bibitem{WTe2QSHEmonolayer_WuS_Science18}
S.~{Wu}, V.~{Fatemi}, Q.~D. {Gibson}, K.~{Watanabe}, T.~{Taniguchi}, R.~J.
  {Cava}, P.~{Jarillo-Herrero}, \emph{Science} \textbf{2018}, \emph{359}, 76.

\bibitem{WTe2SCmonolayer_Fatemi_Science18}
V.~{Fatemi}, S.~{Wu}, Y.~{Cao}, L.~{Bretheau}, Q.~D. {Gibson}, K.~{Watanabe},
  T.~{Taniguchi}, R.~J. {Cava}, P.~{Jarillo-Herrero}, \emph{Science}
  \textbf{2018}, \emph{362}, 926.

\bibitem{WTe2SCmonolayer_Sajadi_Science18}
E.~{Sajadi}, T.~{Palomaki}, Z.~{Fei}, W.~{Zhao}, P.~{Bement}, C.~{Olsen},
  S.~{Luescher}, X.~{Xu}, J.~A. {Folk}, D.~H. {Cobden}, \emph{Science}
  \textbf{2018}, \emph{362}, 922.

\bibitem{LayerSC_Klem_PRB75}
R.~A. Klemm, A.~Luther, M.~R. Beasley, \emph{Phys. Rev. B} \textbf{1975},
  \emph{12}, 877.

\bibitem{MZMsSCsemi_Lutchyn_NRM18}
R.~M. {Lutchyn}, E.~P.~A.~M. {Bakkers}, L.~P. {Kouwenhoven}, P.~{Krogstrup},
  C.~M. {Marcus}, Y.~{Oreg}, \emph{Nat. Rev. Mater.} \textbf{2018}, \emph{3},
  52.

\bibitem{semicSC_Alicea_PRB10}
J.~Alicea, \emph{Phys. Rev. B} \textbf{2010}, \emph{81}, 125318.

\bibitem{SemicSc_Lutchyn_PRL10}
R.~M. {Lutchyn}, J.~D. {Sau}, S.~{Das Sarma}, \emph{Phys. Rev. Lett.}
  \textbf{2010}, \emph{105}, 077001.

\bibitem{SemicSc_Oreg_PRL10}
Y.~Oreg, G.~Refael, F.~von Oppen, \emph{Phys. Rev. Lett.} \textbf{2010},
  \emph{105}, 177002.

\bibitem{Al-InAs_DasA_NP12}
A.~{Das}, Y.~{Ronen}, Y.~{Most}, Y.~{Oreg}, M.~{Heiblum}, H.~{Shtrikman},
  \emph{Nat. Phys.} \textbf{2012}, \emph{8}, 887.

\bibitem{MF2e2/h_LawKT_PRL09}
K.~T. Law, P.~A. Lee, T.~K. Ng, \emph{Phys. Rev. Lett.} \textbf{2009},
  \emph{103}, 237001.

\bibitem{MF2e2/h_LawKT_PRB10}
J.~D. {Sau}, S.~{Tewari}, R.~M. {Lutchyn}, T.~D. {Stanescu}, S.~{Das Sarma},
  \emph{Phys. Rev. B} \textbf{2010}, \emph{82}, 214509.

\bibitem{InSb-NbTiN_Chur_PRB13}
H.~O.~H. Churchill, V.~Fatemi, K.~Grove-Rasmussen, M.~T. Deng, P.~Caroff, H.~Q.
  Xu, C.~M. Marcus, \emph{Phys. Rev. B} \textbf{2013}, \emph{87}, 241401.

\bibitem{Nb-InSb_Deng_NL12}
M.~T. {Deng}, C.~L. {Yu}, G.~Y. {Huang}, M.~{Larsson}, P.~{Caroff}, H.~Q. {Xu},
  \emph{Nano Lett.} \textbf{2012}, \emph{12}, 6414.

\bibitem{InAs-Al_Albre_Nature16}
S.~M. {Albrecht}, A.~P. {Higginbotham}, M.~{Madsen}, F.~{Kuemmeth}, T.~S.
  {Jespersen}, J.~{Nyg{\aa}rd}, P.~{Krogstrup}, C.~M. {Marcus}, \emph{Nature}
  \textbf{2016}, \emph{531}, 206.

\bibitem{KandoEZBP_Gordon_Nat98}
D.~{Goldhaber-Gordon}, H.~{Shtrikman}, D.~{Mahalu}, D.~{Abusch-Magder},
  U.~{Meirav}, M.~A. {Kastner}, \emph{Nature} \textbf{1998}, \emph{391}, 156.

\bibitem{0.7anomalyZBP_Cron_PRL02}
S.~M. Cronenwett, H.~J. Lynch, D.~Goldhaber-Gordon, L.~P. Kouwenhoven, C.~M.
  Marcus, K.~Hirose, N.~S. Wingreen, V.~Umansky, \emph{Phys. Rev. Lett.}
  \textbf{2002}, \emph{88}, 226805.

\bibitem{0.7anomaly_Rokh_PRL06}
L.~P. Rokhinson, L.~N. Pfeiffer, K.~W. West, \emph{Phys. Rev. Lett.}
  \textbf{2006}, \emph{96}, 156602.

\bibitem{InSb-Nb_Rokhinson_NP12}
L.~P. {Rokhinson}, X.~{Liu}, J.~K. {Furdyna}, \emph{Nat. Phys.} \textbf{2012},
  \emph{8}, 795.

\bibitem{Bi2Se3-AlnoMF_Sac_NC11}
B.~{Sac{\'e}p{\'e}}, J.~B. {Oostinga}, J.~{Li}, A.~{Ubaldini}, N.~J.~G.
  {Couto}, E.~{Giannini}, A.~F. {Morpurgo}, \emph{Nat. Commun.} \textbf{2011},
  \emph{2}, 575.

\bibitem{Bi2Te3-PbnoMF_QuF_SR12}
F.~{Qu}, F.~{Yang}, J.~{Shen}, Y.~{Ding}, J.~{Chen}, Z.~{Ji}, G.~{Liu},
  J.~{Fan}, X.~{Jing}, C.~{Yang}, L.~{Lu}, \emph{Sci. Rep.} \textbf{2012},
  \emph{2}, 339.

\bibitem{Be2Se3-Pb_YangF_PRB12}
F.~{Yang}, F.~{Qu}, J.~{Shen}, Y.~{Ding}, J.~{Chen}, Z.~{Ji}, G.~{Liu},
  J.~{Fan}, C.~{Yang}, L.~{Fu}, L.~{Lu}, \emph{Phys. Rev. B} \textbf{2012},
  \emph{86}, 134504.

\bibitem{Bi2Se3-PbNoMF_Williams_PRL12}
J.~R. {Williams}, A.~J. {Bestwick}, P.~{Gallagher}, S.~S. {Hong}, Y.~{Cui},
  A.~S. {Bleich}, J.~G. {Analytis}, I.~R. {Fisher}, D.~{Goldhaber-Gordon},
  \emph{Phys. Rev. Lett.} \textbf{2012}, \emph{109}, 056803.

\bibitem{Bi2Te3-NbNoMF_Veldh_NM12}
M.~{Veldhorst}, M.~{Snelder}, M.~{Hoek}, T.~{Gang}, V.~K. {Guduru}, X.~L.
  {Wang}, U.~{Zeitler}, W.~G. {van der Wiel}, A.~A. {Golubov}, H.~{Hilgenkamp},
  A.~{Brinkman}, \emph{Nat. Mater.} \textbf{2012}, \emph{11}, 417.

\bibitem{Bi2Se3-AlNoMF_ChoS_NC13}
S.~{Cho}, B.~{Dellabetta}, A.~{Yang}, J.~{Schneeloch}, Z.~{Xu}, T.~{Valla},
  G.~{Gu}, M.~J. {Gilbert}, N.~{Mason}, \emph{Nat. Commun.} \textbf{2013},
  \emph{4}, 1689.

\bibitem{HgTe-NbNoMF_Oost_PRX13}
J.~B. {Oostinga}, L.~{Maier}, P.~{Sch{\"u}ffelgen}, D.~{Knott}, C.~{Ames},
  C.~{Br{\"u}ne}, G.~{Tkachov}, H.~{Buhmann}, L.~W. {Molenkamp}, \emph{Phys.
  Rev. X} \textbf{2013}, \emph{3}, 021007.

\bibitem{Bi2Se3-NbNoMF_Finck_PRX14}
A.~D.~K. Finck, C.~Kurter, Y.~S. Hor, D.~J. Van~Harlingen, \emph{Phys. Rev. X}
  \textbf{2014}, \emph{4}, 041022.

\bibitem{QAHE-SC_HeQL_Science17}
Q.~L. {He}, L.~{Pan}, A.~L. {Stern}, E.~C. {Burks}, X.~{Che}, G.~{Yin},
  J.~{Wang}, B.~{Lian}, Q.~{Zhou}, E.~S. {Choi}, K.~{Murata}, X.~{Kou},
  Z.~{Chen}, T.~{Nie}, Q.~{Shao}, Y.~{Fan}, S.-C. {Zhang}, K.~{Liu}, J.~{Xia},
  K.~L. {Wang}, \emph{Science} \textbf{2017}, \emph{357}, 294.

\bibitem{Be2Se3MF_WangE_NP13}
E.~{Wang}, H.~{Ding}, A.~V. {Fedorov}, W.~{Yao}, Z.~{Li}, Y.-F. {Lv},
  K.~{Zhao}, L.-G. {Zhang}, Z.~{Xu}, J.~{Schneeloch}, R.~{Zhong}, S.-H. {Ji},
  L.~{Wang}, K.~{He}, X.~{Ma}, G.~{Gu}, H.~{Yao}, Q.-K. {Xue}, X.~{Chen},
  S.~{Zhou}, \emph{Nat. Phys.} \textbf{2013}, \emph{9}, 621.

\bibitem{Bi2Se3-NbSe2_WangMX_sience12}
M.-X. {Wang}, C.~{Liu}, J.-P. {Xu}, F.~{Yang}, L.~{Miao}, M.-Y. {Yao}, C.~L.
  {Gao}, C.~{Shen}, X.~{Ma}, X.~{Chen}, Z.-A. {Xu}, Y.~{Liu}, S.-C. {Zhang},
  D.~{Qian}, J.-F. {Jia}, Q.-K. {Xue}, \emph{Science} \textbf{2012},
  \emph{336}, 52.

\bibitem{Agdoped(PbSe)5(Bi2Se3)3SC_FangL_PRB14}
L.~Fang, C.~C. Stoumpos, Y.~Jia, A.~Glatz, D.~Y. Chung, H.~Claus, U.~Welp,
  W.-K. Kwok, M.~G. Kanatzidis, \emph{Phys. Rev. B} \textbf{2014}, \emph{90},
  020504.

\bibitem{Misfit_Wiegers_PSSC96}
G.~A. Wiegers, \emph{Prog. Solid State Chem.} \textbf{1996}, \emph{24}, 1.

\bibitem{MisfitHc2_BaiH_JPCM18}
H.~{Bai}, X.~{Yang}, Y.~{Liu}, M.~{Zhang}, M.~{Wang}, Y.~{Li}, J.~{Ma},
  Q.~{Tao}, Y.~{Xie}, G.-H. {Cao}, Z.-A. {Xu}, \emph{J. Phys.-Condes. Matter}
  \textbf{2018}, \emph{30}, 355701.

\bibitem{MoS2SVK_Saito_NP16}
Y.~{Saito}, Y.~{Nakamura}, M.~S. {Bahramy}, Y.~{Kohama}, J.~{Ye},
  Y.~{Kasahara}, Y.~{Nakagawa}, M.~{Onga}, M.~{Tokunaga}, T.~{Nojima},
  Y.~{Yanase}, Y.~{Iwasa}, \emph{Nat. Phys.} \textbf{2016}, \emph{12}, 144.

\bibitem{NbSe22D_XiX_NP16}
X.~{Xi}, Z.~{Wang}, W.~{Zhao}, J.-H. {Park}, K.~T. {Law}, H.~{Berger},
  L.~{Forr{\'o}}, J.~{Shan}, K.~F. {Mak}, \emph{Nat. Phys.} \textbf{2016},
  \emph{12}, 139.

\bibitem{PbSe2DTCI_Wrasse_NL14}
E.~O. {Wrasse}, T.~M. {Schmidt}, \emph{Nano Lett.} \textbf{2014}, \emph{14},
  5717.

\bibitem{IV-VITCI_LiuJ_NanoL15}
J.~{Liu}, X.~{Qian}, L.~{Fu}, \emph{Nano Lett.} \textbf{2015}, \emph{15}, 2657.

\bibitem{SnTe2D_LiuJ_NM14}
J.~{Liu}, T.~H. {Hsieh}, P.~{Wei}, W.~{Duan}, J.~{Moodera}, L.~{Fu}, \emph{Nat.
  Mater.} \textbf{2014}, \emph{13}, 178.

\bibitem{PbS2DTCI_WanWh_AM17}
W.~Wan, Y.~Yao, L.~Sun, C.-C. Liu, F.~Zhang, \emph{Adv. Mater.} \textbf{2017},
  \emph{29}, 1604788.

\bibitem{ZBPKondo_Lee_PRL12}
E.~J.~H. {Lee}, X.~{Jiang}, R.~{Aguado}, G.~{Katsaros}, C.~M. {Lieber}, S.~{De
  Franceschi}, \emph{Phys. Rev. Lett.} \textbf{2012}, \emph{109}, 186802.

\bibitem{ZPBdisorder_LiuJ_PRL12}
J.~Liu, A.~C. Potter, K.~T. Law, P.~A. Lee, \emph{Phys. Rev. Lett.}
  \textbf{2012}, \emph{109}, 267002.

\bibitem{ZBPkondo_Pikulin_NPJ12}
D.~I. {Pikulin}, J.~P. {Dahlhaus}, M.~{Wimmer}, H.~{Schomerus}, C.~W.~J.
  {Beenakker}, \emph{New J. Phys.} \textbf{2012}, \emph{14}, 125011.

\bibitem{ZBPkondo_Kells_PRB12}
G.~Kells, D.~Meidan, P.~W. Brouwer, \emph{Phys. Rev. B} \textbf{2012},
  \emph{86}, 100503.

\bibitem{Fe-PbZBP_Nadj_Science14}
S.~{Nadj-Perge}, I.~K. {Drozdov}, J.~{Li}, H.~{Chen}, S.~{Jeon}, J.~{Seo},
  A.~H. {MacDonald}, B.~A. {Bernevig}, A.~{Yazdani}, \emph{Science}
  \textbf{2014}, \emph{346}, 602.

\bibitem{Zeromode_Clarke_NatP14}
D.~J. {Clarke}, J.~{Alicea}, K.~{Shtengel}, \emph{Nat. Phys.} \textbf{2014},
  \emph{10}, 877.

\bibitem{MZMinGraphene_SanJP_PRX15}
P.~{San-Jose}, J.~L. {Lado}, R.~{Aguado}, F.~{Guinea},
  J.~{Fern{\'a}ndez-Rossier}, \emph{Phys. Rev. X} \textbf{2015}, \emph{5},
  041042.

\bibitem{FeSeTeTSC_ZhangP_Science18}
P.~{Zhang}, K.~{Yaji}, T.~{Hashimoto}, Y.~{Ota}, T.~{Kondo}, K.~{Okazaki},
  Z.~{Wang}, J.~{Wen}, G.~D. {Gu}, H.~{Ding}, S.~{Shin}, \emph{Science}
  \textbf{2018}, \emph{360}, 182.

\bibitem{PbTaSe2_BianG_NatC}
G.~Bian, T.-R. Chang, R.~Sankar, S.-Y. Xu, H.~Zheng, T.~Neupert, C.-K. Chiu,
  S.-M. Huang, G.~Q. Chang, I.~Belopolski, D.~S. Sanchez, M.~Neupane,
  N.~Alidoust, C.~Liua, B.~K. Wang, C.-C. Lee, H.-T. Jeng, C.~L. Zhang, Z.~J.
  Yuan, S.~Jia, A.~Bansil, F.~C. Chou, H.~Lin, M.~Z. Hasan, \emph{Nat. Commun.}
  \textbf{2016}, \emph{7}, 10556.

\bibitem{PbTaSe2_GuanSY_SciAdv16}
S.-Y. {Guan}, P.-J. {Chen}, M.-W. {Chu}, R.~{Sankar}, F.~{Chou}, H.-T. {Jeng},
  C.-S. {Chang}, T.-M. {Chuang}, \emph{Sci. Adv.} \textbf{2016}, \emph{2},
  e1600894.

\bibitem{CatalogueTM_ZhangTT_nature19}
T.~{Zhang}, Y.~{Jiang}, Z.~{Song}, H.~{Huang}, Y.~{He}, Z.~{Fang}, H.~{Weng},
  C.~{Fang}, \emph{Nature} \textbf{2019}, \emph{566}, 475.

\bibitem{TM_TangF_nature19}
F.~{Tang}, H.~C. {Po}, A.~{Vishwanath}, X.~{Wan}, \emph{Nature} \textbf{2019},
  \emph{566}, 486.

\bibitem{TM_VergnioryMG_nature19}
M.~G. {Vergniory}, L.~{Elcoro}, C.~{Felser}, B.~A. {Bernevig}, Z.~{Wang},
  \emph{Nature} \textbf{2019}, \emph{566}, 480.

\end{thebibliography}

\end{document}